\documentclass[12pt,preprint]{aastex}

\newcommand{\kms}{\mbox{\,km\,s$^{-1}$}}
\newcommand{\Kkms}{\mbox{\,K\,km\,s$^{-1}$}}
\newcommand{\Msun}{\,$M_{\odot}$}
\newcommand{\Mj}{\,$M_{J}$}

\newcommand{\co}{\mbox{\rmfamily $^{12}$CO}}
\newcommand{\tco}{\mbox{\rmfamily $^{13}$CO}}
\newcommand{\ceo}{\mbox{\rmfamily C$^{18}$O}}
\newcommand{\ceol}{\mbox{\rmfamily C$^{18}$O\,}{(1--0)}}

\newcommand{\ceoh}{\mbox{\rmfamily C$^{18}$O\,}{(2--1)}}

\newcommand{\vlsr}{v$_{LSR}$}
\newcommand{\degree}{$^{\circ}$}
\newcommand{\cms}{${\rm cm}^{-2}$}
\newcommand{\cmc}{${\rm cm}^{-3}$}
\newcommand{\hh}{H$_{2}$}
\newcommand{\hhco}{H$_{2}$CO}
\newcommand{\tastar}{T$_{A}^{*}$}
\newcommand{\tmb}{T$_{mb}$}

\newcommand{\tk}{T$_{K}$}
\newcommand{\av}{A$_{V}$}
\newcommand{\nnhp}{N$_{2}$H$^{+}$}
\newcommand{\ammonia}{NH$_{3}$}

\slugcomment{}
\shorttitle{Coalsack: the merging of two subsonic flows}
\shortauthors{Rathborne et al.}

\begin{document}

\title{The dense ring in the Coalsack: the merging of two subsonic flows?}
\author{J. M. Rathborne,  C. J. Lada and W. Walsh}
\affil{Harvard-Smithsonian Center for Astrophysics, 60 Garden Street, Cambridge, MA 02138, USA: jrathborne@cfa.harvard.edu, clada@cfa.harvard.edu, wwalsh@cfa.harvard.edu}
\and
\author{M. Saul}
\affil{School of Physics, The University of New South Wales, Sydney, NSW 2052, Australia: msaul@phys.unsw.edu.au}
\and
\author{H. M. Butner}
\affil{Department of Physics and Astronomy, James Madison University, Harrisonburg, VA 22807, USA: butnerhm@jmu.edu}
\begin{abstract}
A recent high angular resolution extinction map toward the most opaque
molecular globule, Globule 2, in the Coalsack Nebula revealed that it
contains a strong central ring of dust column density. This ring
represents a region of high density and pressure that is likely a
transient and possibly turbulent structure. Dynamical models suggest
that the ring has formed as a result of a sudden increase in external
pressure which is driving a compression wave into the Globule. Here we
combine the extinction measurements with a detailed study of the
\ceol\, molecular line profiles toward Globule 2 in order to
investigate the overall kinematics and, in doing so, test this
dynamical model.  We find that the ring corresponds to an enhancement
in the \ceo\, non-thermal velocity dispersion and non-thermal
pressure. We observe a velocity gradient across the Globule that appears to
trace two distinct systematic subsonic velocity flows that happen
to converge within the ring. We suggest, therefore, that the ring has
formed as two subsonic flows of turbulent gas merge within the
Globule. The fact that the outer layers of the Globule appear stable
against collapse yet there is no centrally condensed core, suggests
that the Globule may be evolving from the outside in and has yet to
stabilize, confirming its youth.
\end{abstract}


\section{Introduction}

Understanding the initial conditions that lead to the formation of
dense cores is important as it has direct implications for theories of
star formation.  Recent studies of starless cores reveal that the
kinematics in the early stages of core collapse can be quite complex:
in some cases the molecular tracers indicate simultaneously opposing
motions (expansion {\it {and}} infall; \citealp{Lada03a}). In the case
of the starless Bok globule B68, the presence of both inward and
outward motions, traced by molecular line emission, has been
interpreted to arise due to the oscillation of the outer layers of the
cloud \citep{Lada03a,Redman06,Maret07}. This idea is strengthened by modeling
that suggests that the complex motions within starless cores can occur due
to oscillating perturbations about a thermally-supported, pressure
bounded structure \citep{Keto06,Broderick07}. Indeed, recent numerical
simulations reveal that small amounts of rotation can cause initially
unstable and collapsing cores to stabilize and oscillate
\citep{Matsumoto03}.

The combination of high angular resolution extinction and molecular
line observations toward cold, dense starless cores can trace the
properties and kinematics of the dense gas and allow us to gain
insight into the physical nature of core and, thus, star
formation. Extinction measurements are particularly useful for tracing
the internal structure of cores because they probe the material along
the line-of-sight and are not susceptible to optical depth effects (as
in the case of \co\, molecular line emission) or assumptions about the
dust temperature or emissivity (as in the case of millimeter continuum
emission). Thus, the combination of dust extinction observations with
optically thin molecular line observations can accurately reveal the
masses and densities of the dust as well as the kinematics and
dynamics of the gas.

In this paper we combine extinction observations with a detailed
kinematic study of the molecular line profiles observed toward Tapia's
Globule 2 in the Coalsack Nebula. Because Globule 2 does not contain a
central protostar, is one of the densest cores within the Coalsack
complex, and is thought to be in a very early stage of condensing, a
detailed study of its kinematic structure could provide important
details on the processes leading to core formation.

The Coalsack Nebula is a nearby dark molecular cloud complex that
extends almost 6\degree\, on the sky ($\sim$ 15 pc at its distance of
150 pc; \citealp{Cambresy99}) and appears completely devoid of star
formation activity \citep{Nyman89,Bourke95,Kato99}. It has a total
mass of $\sim$ 3550\,\Msun\, and is characterized by complex,
filamentary molecular structures which contain many dark cores with
kinetic temperatures, \tk, of 8--11 K \citep{Nyman89}. The darkest and
densest of these cores are Tapia's Globules 1, 2, and 3
\citep{Tapia73,Bok77}.

Using deep near-infrared imaging observations obtained with the
European Southern Observatory (ESO) 3.6m New Technology Telescope
(NTT), \cite{Lada04} were able to examine the structure of Globule 2
in significantly greater detail than previously possible. The NICE
method \citep{Lada94,Alves98} was used to generate a visual extinction (\av) 
map that was constructed by measuring the individual infrared extinctions to
approximately 24,000 stars located behind Globule 2.  This resulted in
an extinction map across the core with significantly improved angular
resolution ($\sim$ 15\arcsec) compared to previous maps
($\sim$ 60\,\arcsec; \citealp{Jones80,Jones84,Racca02}).

The extinction map revealed Globule 2 to have a mass of
$\sim$6.1\,\Msun, a mean density of $\sim$3$\times$10$^{3}$\,\cmc\,
and Jeans mass, \Mj, of $\sim$10.4\,\Msun. From near-infrared
polarimetry observations, \cite{Jones84} suggested that Globule 2 is
supported either by turbulence or a magnetic field and appears
stable. This is also consistent with the fact that the derived mass is
less than \Mj\, suggesting it is unlikely to fragment
further. Moreover, a Bonnor-Ebert (BE) fit to the outer regions of the
Globule's radial extinction profile suggests that it is globally
stable against collapse \citep{Lada04}. The Globule's derived BE mass
is $\sim$ 6\,\Msun. Even though this is comparable to the derived mass
calculated from the dust, the high angular resolution infrared
extinction map toward the Globule does not reveal a centrally
condensed core as would be expected if it was a true BE sphere.
Instead, the most prominent feature within the extinction map is a
strong central ring of dust column density. This ring likely
represents a transient structure that has formed from a region of high
density and pressure
\citep{Lada04}. Because there is no source of internal pressure to
support the ring against gravitational collapse, it is thought that
the ring is not in dynamical equilibrium with its surroundings and,
given the short timescales for inward collapse (2$\times$10$^5$
years), most likely represents an extremely early stage of evolution.

From a single \ceoh\, spectrum toward the center of the globule,
\cite{Lada04} noted a double-peaked profile comprising 
a bright narrow line and a fainter broad line. Based on the absence of
this double-lined profile in lower angular resolution spectra
\citep[e.g.,][]{Brooks76,Nyman89,Kato99}, it was suggested
that the narrow line component most likely arises within a small
region and that the broad line component traces a more spatially
extended feature \citep{Lada04}. The line width of each blended
component was found to be super-thermal indicating that significant
non-thermal, or turbulent, motions most likely dominate.  Dynamical
models for Globule 2, however, suggest that the ring structure has
formed as a result of a sudden increase in external pressure which is
driving a compression wave into the Globule \citep{Hennebelle06}.

Using the Mopra 22-m telescope, we have obtained the first complete
high angular and spectral resolution molecular line map of Globule 2.
Because \co\, is optically thick toward most dense cores, and is thus
only tracing the outer layers, we choose to conduct this study using
the optically thin \ceol\, transition which is far superior in tracing
the dense gas associated with the central core.  In combination with
the extinction map, these data should allow us to test the various
formation mechanisms for the ring feature; is it a high-density
transient structure formed via the dissipation of turbulent gas
\citep{Lada04} or does it trace the propagation of a compression wave 
through the Globule \citep{Hennebelle06}?  Moreover, these data will
enable the relative spatial distributions of the two molecular line
components to be definitively traced.  In particular, does the
narrow-line component originate exclusively in the dense ring?  Does
the broad-line component primarily trace the turbulent envelope of the
Globule?  Answers to these questions will directly test the idea that
Globule 2 is in perhaps the earliest stage of evolution: a dense core
in the process of formation.

\section{Observations}

Molecular line maps of Globule 2 in the Coalsack Nebula were obtained
with the Mopra 22-m telescope in 2006 July. The 8\,GHz spectrometer
MOPS was used to simultaneously observe a number of transitions
including \ceol\,\footnote{The other transitions observed included
\co, \tco, HCO$^{+}$, CH$_{3}$OH, and CS and will be presented in
future papers (Saul et al., Butner et al. in prep.).}.  The
spectrometer was used in `zoom' mode such that one spectral window
covered each of the lines of interest. Each window was 137.5 MHz wide
and contained 4096 channels in both orthogonal polarizations.  This
produced a velocity resolution of 0.09\,\kms\, for the \ceo\,
transition. At these frequencies the Mopra beam is $\sim$ 33\arcsec\,
\citep{Ladd05}.

The observations were obtained using the On-The-Fly (OTF) mode by
scanning in both Right Ascension and Declination. All maps were
Nyquist sampled.  A fixed off-source position was used for all maps
($\alpha$=12:29:57.1,$\delta$=$-$63:32:21.8, J2000).  The nearby SiO
maser, R Car, was used to check the pointing accuracy approximately
every hour and was typically better than 9\arcsec. The system
temperatures were checked frequently using a paddle and were typically
180\,K.

A total of fourteen 5\arcmin$\times$5\arcmin\, OTF maps were combined
(using the system temperature weighted median) to produce the final
map.  The data were initially baseline subtracted and converted to a
temperature scale using the AIPS$^{++}$ task livedata. All raw data
are in the \tastar\, scale. To convert to the main beam brightness,
\tmb, we use the beam efficiencies listed in \cite{Ladd05}.

The OTF maps were combined and gridded into data cubes using the
AIPS$^{++}$ task gridzilla. To produce the final map, a Gaussian
smoothing kernel with a FWHM of 0.55\arcmin\, was applied
to the data. This resulted in the final cube having an effective
angular resolution of $\sim$ 46\arcsec. The final
\ceo\, map has a \tastar\,  rms noise sensitivity of 0.08 K channel$^{-1}$.

\section{Results}

\begin{figure}
\begin{center}
\includegraphics[angle=-90,width=0.7\textwidth]{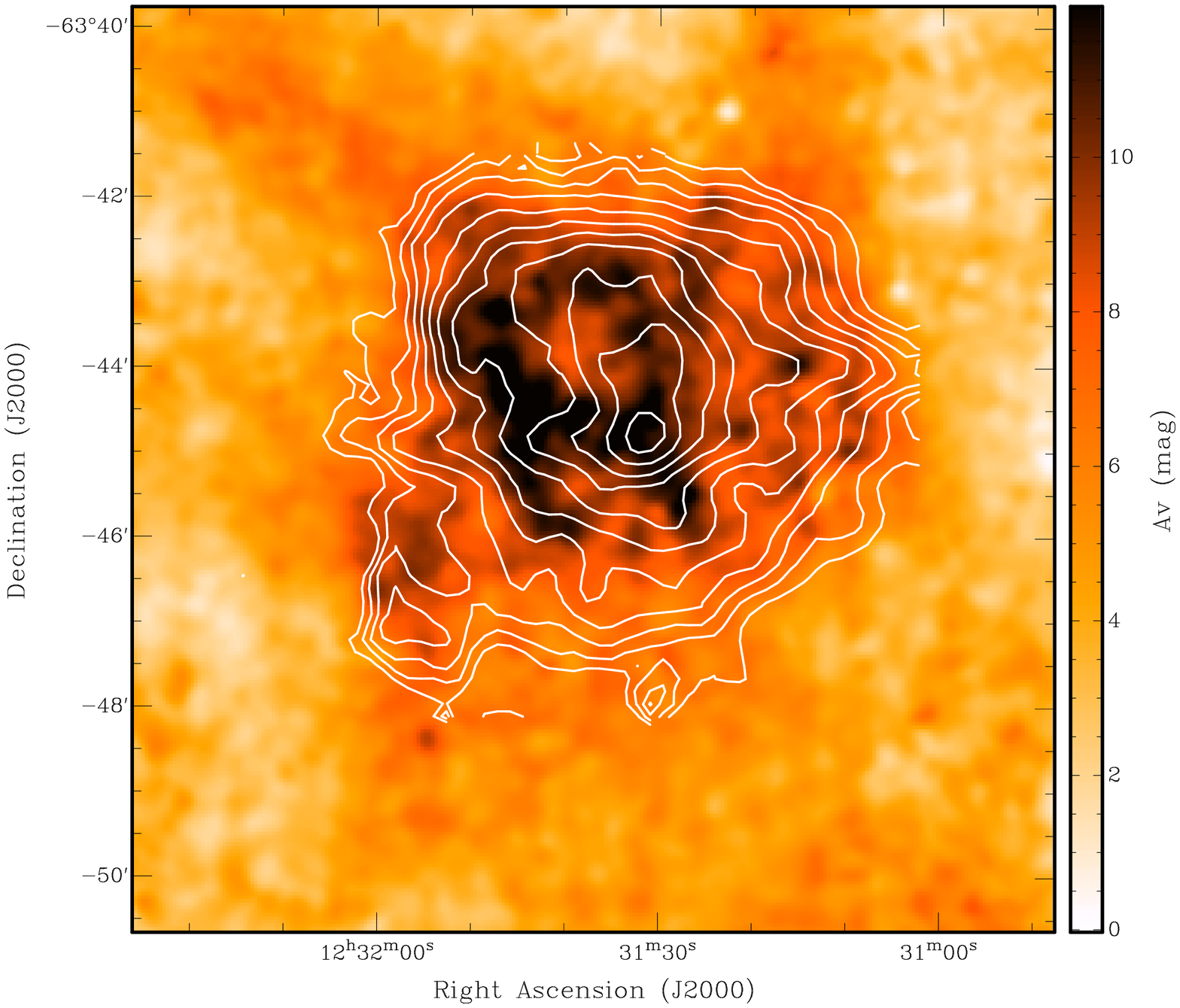}
\caption{\label{av-c18o}The visual extinction (\av) map of Globule 2 in the 
    Coalsack Nebula (15\arcsec\, angular resolution; \citealp{Lada04})
    in color scale overlaid with contours of the \ceo\, integrated
    intensity (46\arcsec\, angular resolution). The emission was
    integrated over the velocity range of $-$7.0 to $-$5.0\,\kms. The
    contour levels are from 1.0 to 2.0 in steps of 0.1 \Kkms. Note
    that the general morphology of the \ceo\, integrated intensity
    follows the enhanced extinction associated with the ring.}
\end{center}
\end{figure}

Figure~\ref{av-c18o} shows the visual extinction map of Globule 2
(15\arcsec\, angular resolution).  This map was generated from deep
near-infrared images by measuring the extinction toward $\sim$ 24,000
background stars \citep{Lada04}. The ring is clearly seen as an
extinction enhancement surrounding a central depression. Overlaid on
this map is the \ceol\, integrated emission obtained with the Mopra
22-m telescope (the emission was integrated over the velocity range of
$-$7.0 to $-$5.0 \kms; 46\arcsec\, angular resolution).  These data
show that the \ceo\, integrated emission follows the overall
morphology of the extinction. There are noticeable differences,
however. The \ceo\, integrated emission does not trace completely the
extinction ring.  In particular, it appears that the extinction peaks
toward the eastern arc of the ring, whereas the \ceo\, integrated
emission peaks toward western arc. Some of these differences may arise
due to the different angular resolutions of the two datasets
(15\arcsec\, and 45\arcsec\, respectively). When smoothed to an
angular resolution of 45\arcsec, the extinction image shows a much
shallower contrast between the ring and the central hole.

\begin{figure}
\begin{center}
\includegraphics[angle=-90,width=0.9\textwidth]{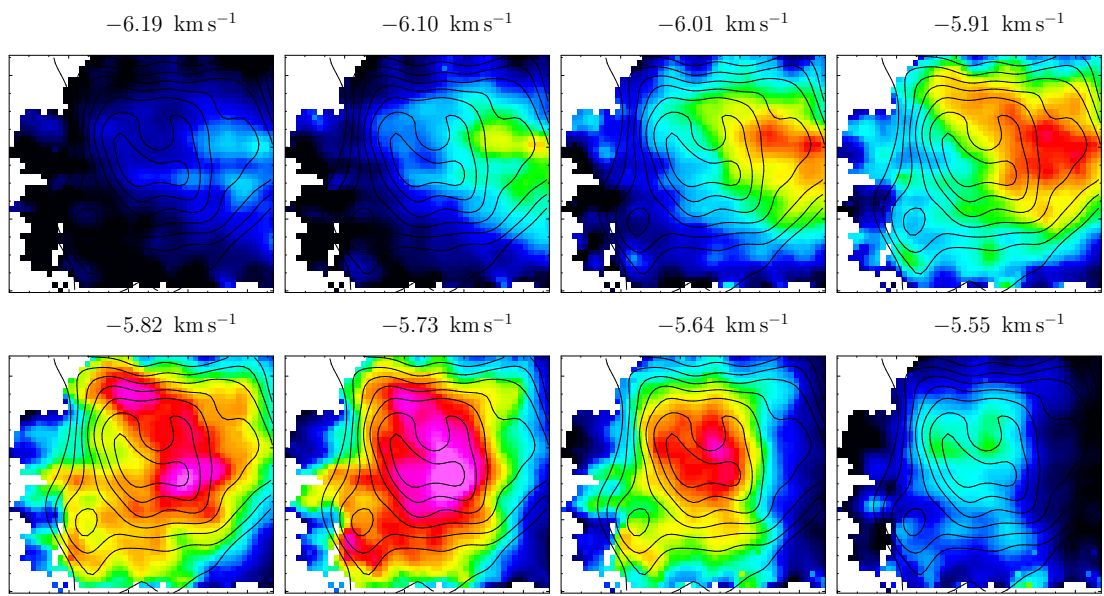}
\caption{\label{channels}\ceol\, channel maps overlaid with contours of 
    the smoothed visual extinction map toward Globule 2 (46\arcsec\,
    angular resolution). The contour levels are from 6 to 12 in steps
    of 0.75 magnitudes. The color maps show a single channel centered
    at the \vlsr\, marked above each image. All images cover the same
    extent in Right Ascension and Declination ($\sim$ 6\arcmin\,
    $\times$ 6\arcmin) and have the same color scale from a \tmb\,
    of 0 (black) to 3.4 K (pink). We see a clear velocity gradient
    across the Globule.}
\end{center}
\end{figure}

Figure~\ref{channels} shows the \ceol\, channel maps across the
Globule overlaid with contours of the smoothed visual extinction image
(the \av\, map of Fig.~\ref{av-c18o} was smoothed to 46\arcsec\, to
match the angular resolution of the \ceo\, data). A clear velocity
gradient is evident across the Globule.

\subsection{Gaussian component analysis}

The individual \ceol\, spectra across the Globule reveal that the
emission profile differs quite considerably on and off the extinction
ring. Based on the detection of a `narrow-line'
(\vlsr$\sim$$-$5.64\,\kms) and `broad-line' (\vlsr$\sim$$-$5.83\,\kms)
component within the \ceo\,(2--1) spectrum toward the ring and the
absence of the narrow-line component within lower-angular resolution
surveys, it was suggested that the broad line was spatially extended
across the Globule while the narrow line was restricted to the
extinction ring \citep{Lada04}. Thus, to characterize and separate
these two emission features and determine any spatial differences
between them, we fit Gaussian profiles to each spectrum within the
map.

Because we expect the broad-line component to contribute to the
observed spectra at all points across the map, but the narrow-line
component to only those spectra coincident with the extinction ring,
we attempt to fit both a one- and two-component Gaussian profile to
each spectrum within the map. Initial guesses for the \vlsr, line
width, and peak temperature for the Gaussian profiles were based on
the observed profiles from \cite{Lada04}. All were free
parameters\footnote{the Gaussian fitting was performed within IDL
using the gaussfits procedure.}. For example, when fitting a
one-component Gaussian profile we use as input the values determined
for the `broad' line. When fitting a two-component Gaussian profile we
input the values derived for both the `broad' and `narrow' lines.  At
each position within the map we select either the one- or
two-component Gaussian fit based on which had the lower residuals. We
then assigned each fitted Gaussian profile to either the `narrow' or
`broad' line component based on its derived \vlsr. We find that $\sim$
70\% of the spectra were well fit by a one-component Gaussian profile,
while the remaining $\sim$ 30\% were best fit by two components.

As expected, we find that the spectra toward the extinction ring are
best matched by a two-component Gaussian profile, while the spectra
external to the ring are best matched by a single component.  Contrary
to the original prediction, however, the central \vlsr\, derived for
the emission external to the ring does not correspond to a single
\vlsr, nor does it simply correspond to the `broad' component.
Instead, the \vlsr\, shows a range of values (from $-$6.0 to
$-$5.7\,\kms) which suggests that both the `broad' and `narrow' lines
are located external to the ring. It appears, therefore, that the
narrow-line component may not simply arise within the ring, but
instead may correspond to a spatially separate velocity component
extended across the Globule.

\begin{figure}
\begin{center}
\includegraphics[angle=90,width=0.7\textwidth]{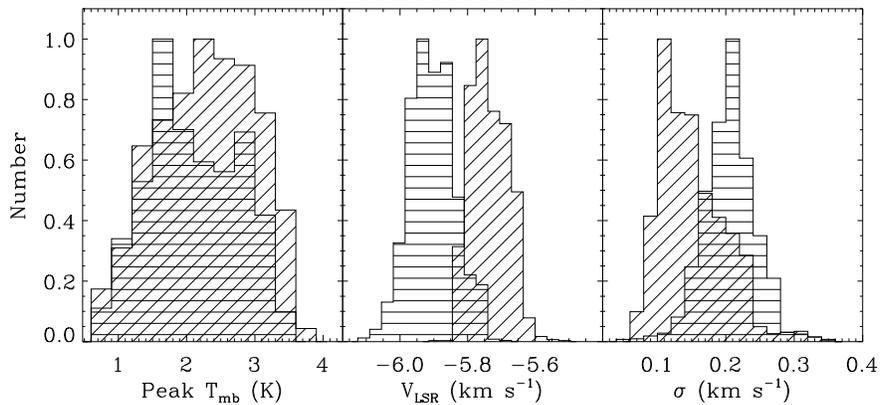}
\caption{\label{gauss-fits}Normalized number distributions for the peak temperature 
         (\tmb), central velocity (\vlsr), and one-dimensional
         velocity dispersion ($\sigma$) output from the Gaussian
         fitting procedure. In all cases the histograms with
         horizontal lines represent the data for the `broad' line
         component, while histograms with diagonal lines represent the
         `narrow' line component. Note that the two components show a
         range in their central \vlsr, suggesting the kinematics of
         the Globule are more complicated than two simple Gaussian
         emission profiles.}
\end{center}
\end{figure}

This can be seen in the \ceo\, channel maps of Figure~\ref{channels}:
the component at $\sim$$-$5.64\,\kms\, peaks in the ring, but also
extends beyond it. The broad-line component ($\sim$$-$5.82\,\kms) also
arises both in and out of the ring, the only difference being it
appears to trace the molecular material on the opposing side of the
extinction ring to the narrow-line component.  Figure~\ref{gauss-fits}
shows the measured peak temperature (\tmb), central velocity (\vlsr),
and one-dimensional velocity dispersion ($\sigma$) for the two
components from the Gaussian fitting procedure. We find that the
individual measured central \vlsr\, and $\sigma$ for the two
components show a dispersion in their values.  This suggests that the
kinematics of the region are quite complex and not well matched simply
by two Gaussian \vlsr\, components toward the ring and a single
\vlsr\, component external to it.

The absence of the narrow-line component in previous molecular line
surveys is not surprising given their limited angular and velocity
resolutions. Taking an average \ceol\, spectra over the region
contained within the beam of the \hhco\, observations (4\farcm4;
\citealp{Brooks76}), we find that the molecular line profile is
characterized by a single emission line, with \vlsr $\sim$
$-$5.8\,\kms\, and line width, $\Delta$V, of 0.6\,\kms.

\subsection{Moment analysis}

Rather than refer to the individual components as the `narrow' and
`broad' lines based on their original \vlsr\, and Gaussian
identification, these data reveal that the \ceo\, emission displays
complex velocity features across the Globule.  Thus, in this section
we characterize the emission via a moment analysis which is
independent of Gaussian deconvolution and assumptions about individual
components.

\begin{figure}
\begin{center}
\includegraphics[angle=-90,totalheight=0.13\textheight,clip=true]{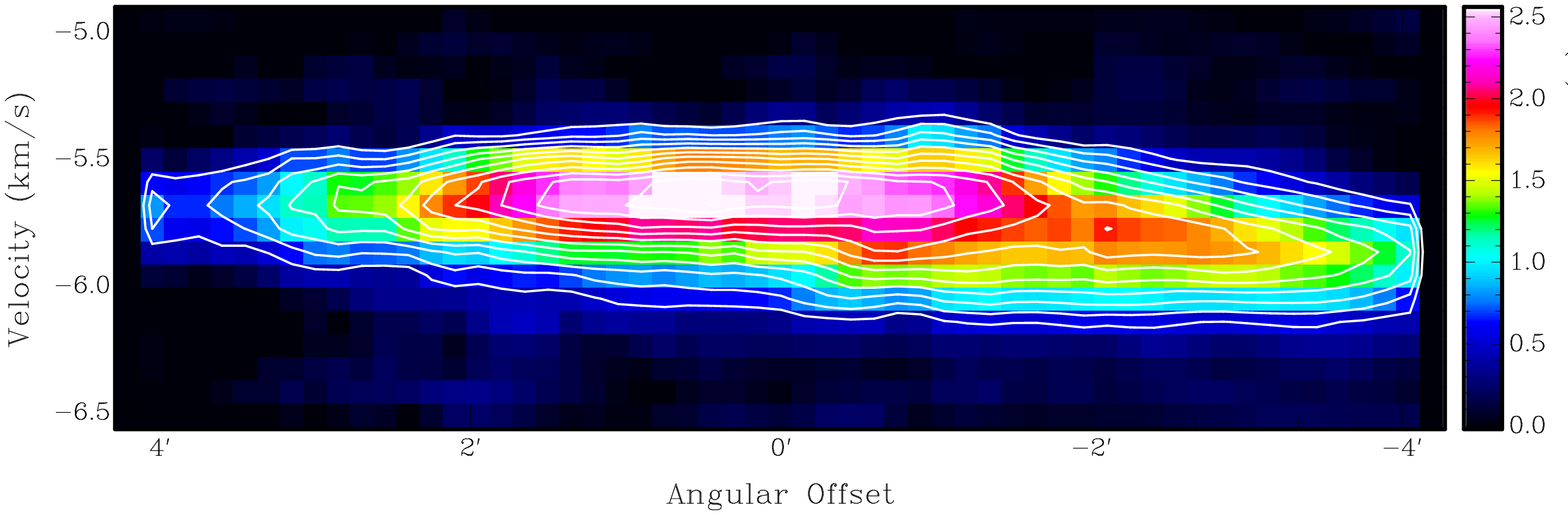}
\includegraphics[angle=-90,totalheight=0.13\textheight,clip=true]{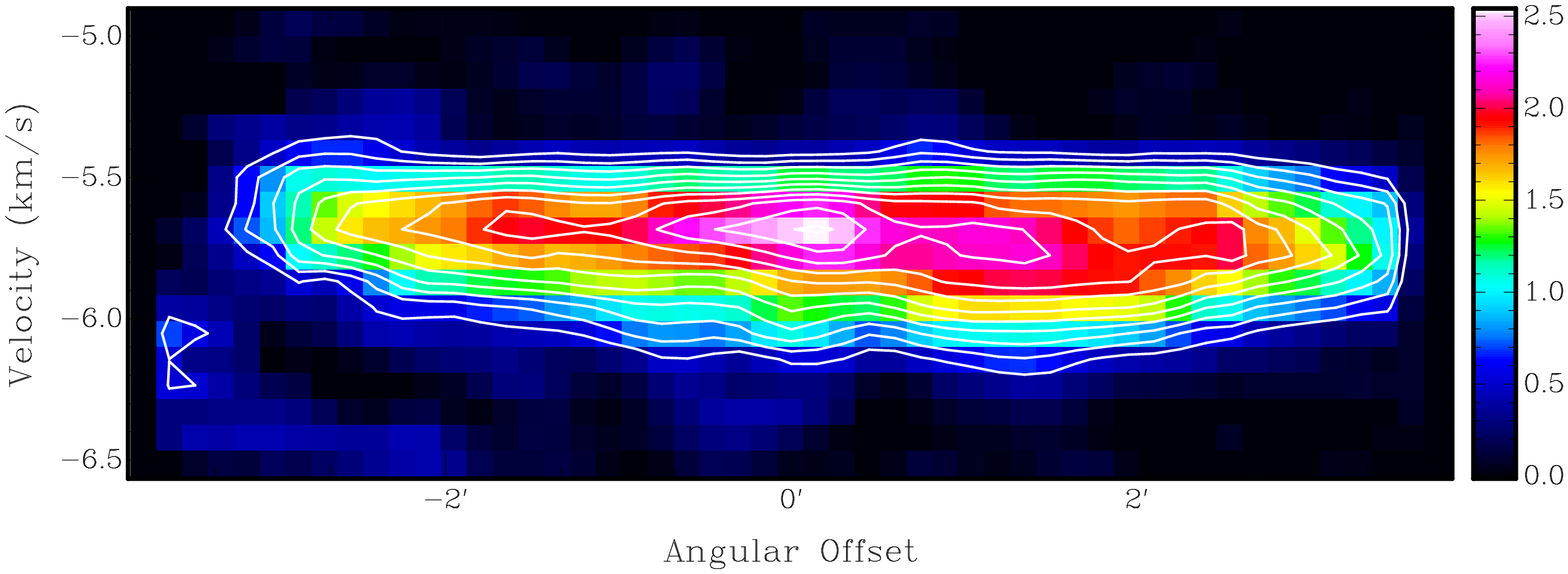}
\caption{\label{c18o-lv}\ceol\, position-velocity, ($\ell$-$V$), diagram (color-scale 
   and contours). The emission was averaged over the extinction ring
   in both Declination (left) and Right Ascension (right). A velocity
   gradient is most apparent in the left panel (the
   Declination-averaged position-velocity diagram) which is not
   surprising given this direction is roughly perpendicular to the
   axis of the apparent gradient in the first moment map (see
   Fig.~\ref{c18o-mom}).}
\end{center}
\end{figure}

Over the extent of the Globule, the combination of these velocity
features manifest themselves as a velocity gradient.  This velocity
gradient is seen in the \ceo\, channel maps
(Fig.~\ref{channels}). While the brightest emission coincides with the
extinction ring, we also see the red- and blue-shifted emission
tending to arise on opposite sides of the extinction ring.  This can
also be seen in the position-velocity, ($\ell$-$V$), diagrams of
Figure~\ref{c18o-lv}. The left panel shows the ($\ell$-$V$) diagram
averaged over the extinction ring in Declination (the emission was
averaged over $\Delta \delta$ of 6\arcmin\, centered on the ring). The
right panel shows the Right Ascension-averaged diagram (the emission
was averaged over $\Delta
\alpha$ of 18\arcmin\, centered on the ring). The velocity gradient is
more apparent in the Declination-averaged ($\ell$-$V$) diagram.

\begin{figure}
\begin{center}
\includegraphics[angle=-90,width=0.47\textwidth]{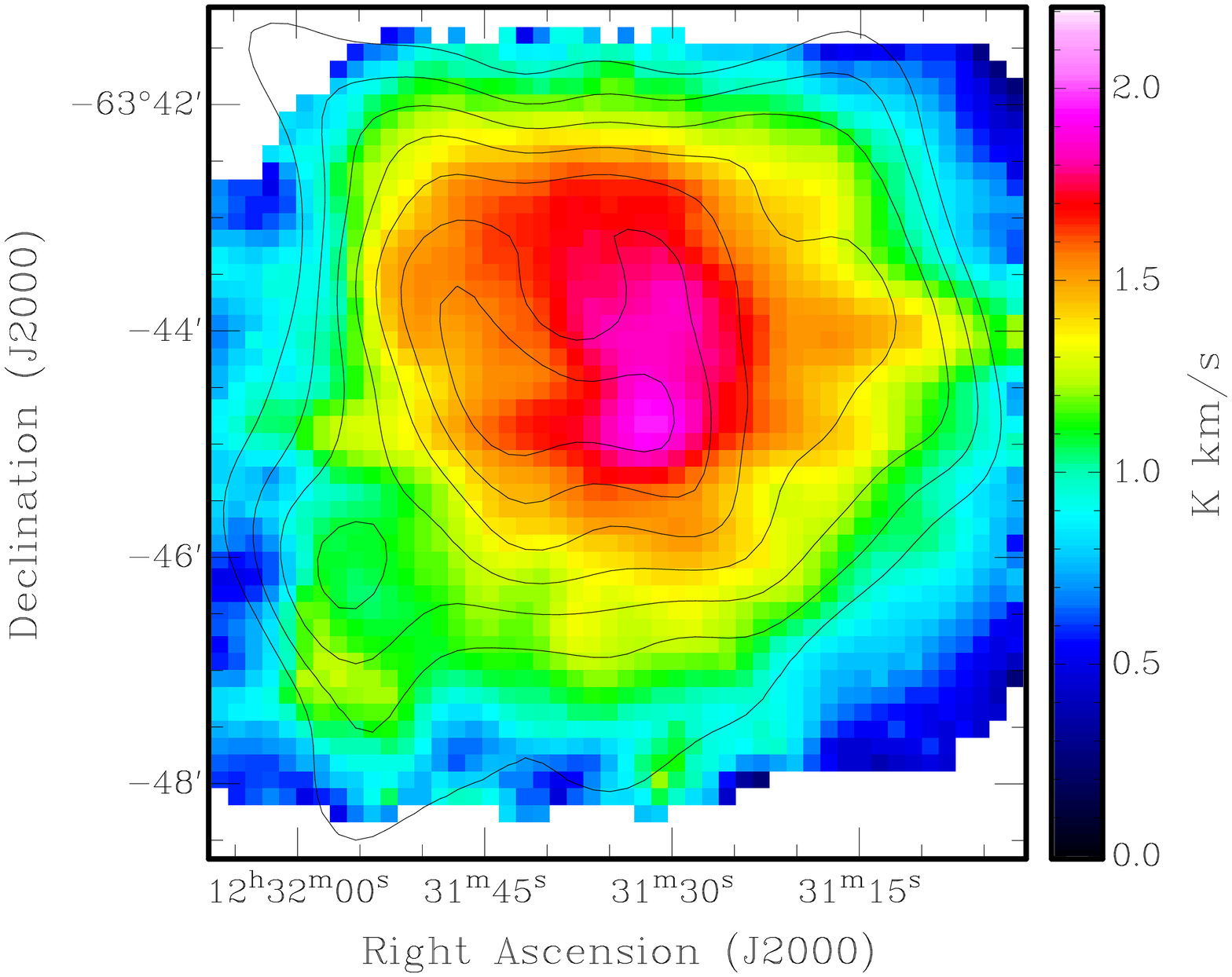}
\includegraphics[angle=-90,width=0.47\textwidth]{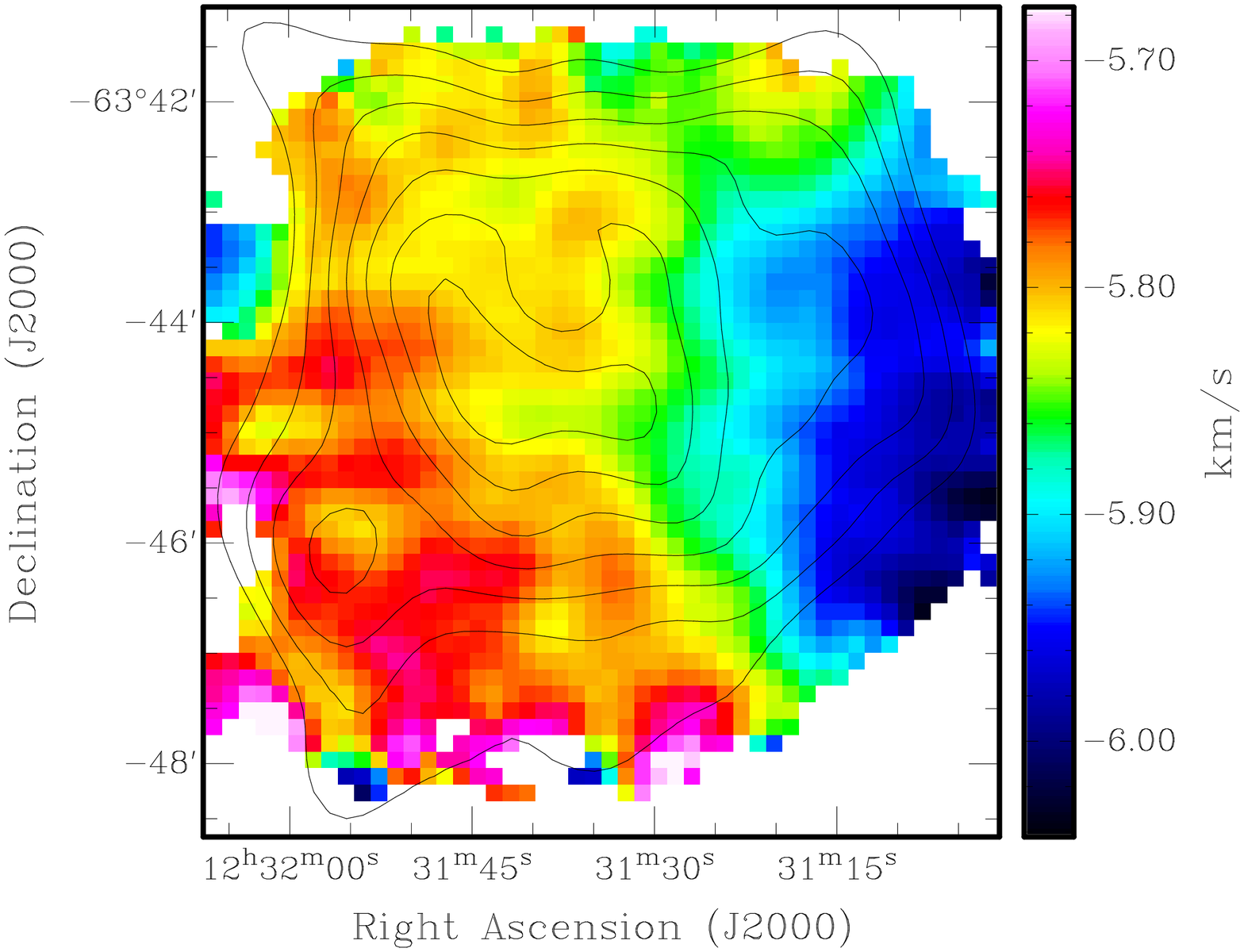}\\
\includegraphics[angle=-90,width=0.47\textwidth]{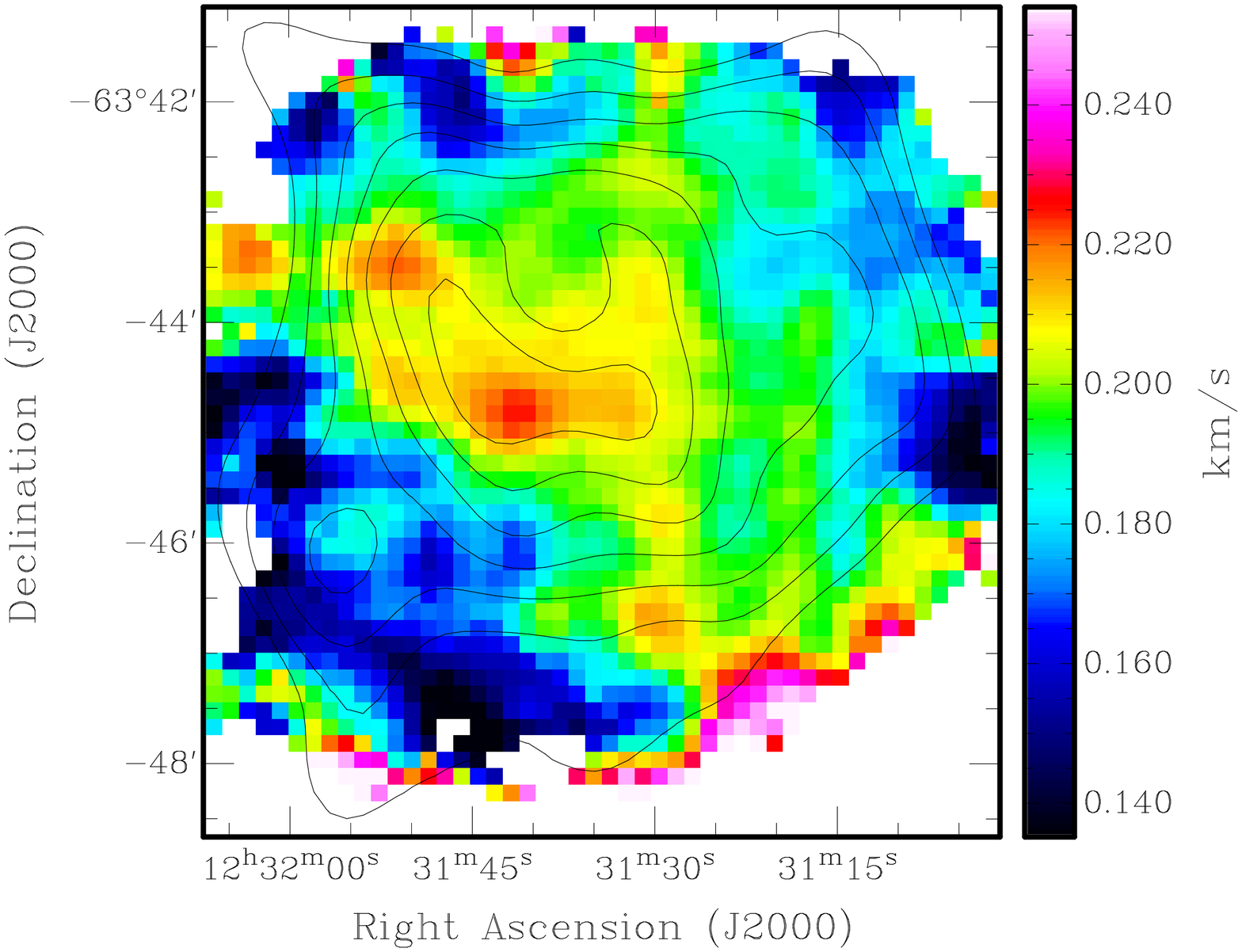}
\includegraphics[angle=-90,width=0.47\textwidth]{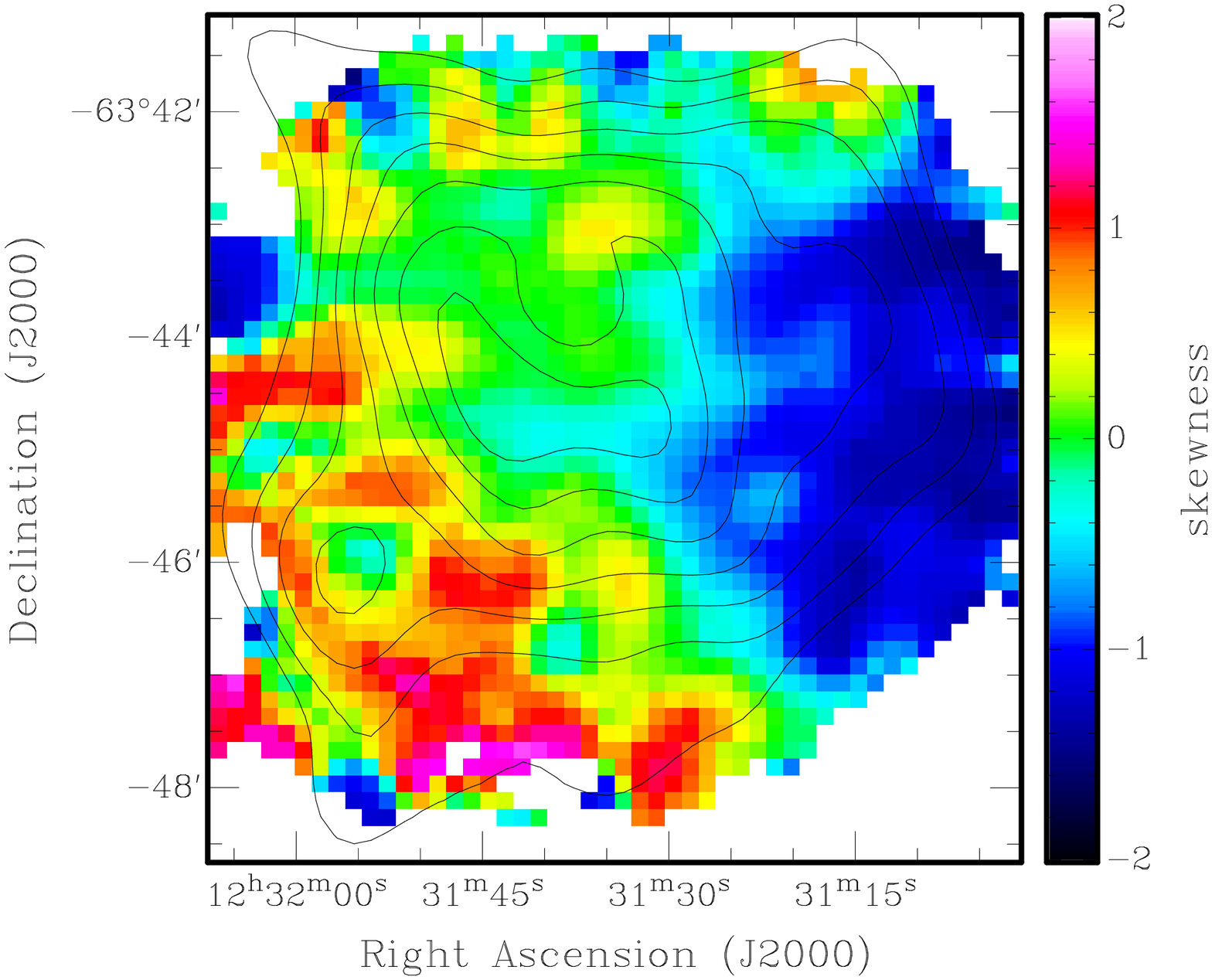}\\
\caption{\label{c18o-mom}\ceol\, moment maps in color scale with contours of the smoothed extinction
   image. {\it {Top left}}: Zeroth moment (integrated intensity). {\it
   {Top right}}: First moment (intensity weighted velocity
   field). {\it {Bottom left}}: Second moment (velocity
   dispersion). {\it {Bottom right}}: Skewness image.  The contour
   levels are from 6 to 12 in steps of 0.75 magnitudes. The extinction
   ring is seen as an enhancement in the \ceo\, integrated
   intensity. The first moment map reveals a velocity gradient across
   the Globule. The velocity dispersion measured by the second moment
   map clearly increases toward the extinction ring. The skewness
   image reveals that the profiles change from a negative (blue)
   asymmetry to a positive (red) asymmetry across the Globule.}
\end{center}
\end{figure}

The zeroth moment map, or integrated intensity image, is defined as

\[ I = \int T_{mb}(v)\,dv \] 

\noindent where \tmb$(v)$ is the main beam brightness at a given velocity
$v$. The image was generated by integrating all of the emission over
the range of $-$7.0 to $-$5.0\,\kms\, and is shown in
Figure~\ref{c18o-mom} overlaid with contours of the smoothed visual
extinction image. The extinction ring is clearly seen as an
enhancement in \ceo\, emission.

The first moment map, M$_{1}$, is a measure of the intensity (in this
case \tmb) weighted velocity field and is defined as
\[M_{1} = \frac{\int T_{mb}(v)v\, dv}{\int T_{mb}(v)\,dv} \]
\noindent \citep{Sault95}. The \ceo\, first moment map is
also shown in Figure~\ref{c18o-mom} and clearly shows the
aforementioned velocity gradient. Toward the edges of the extinction
ring the velocity field ranges from $-$6.0 to $-$5.7\,\kms\, Toward
the peak of the extinction ring, the \vlsr\, of the emission is
$\sim$~$-$5.8\,\kms.

The intensity weighted velocity dispersion is represented via the
second moment, M$_{2}$, which is defined as

\[ M_{2} = \sqrt{\frac{\int T_{mb}(v) (v-M_{1})^{2}\,dv}{\int T_{mb}(v)\,dv}} \]

\noindent The \ceo\, second moment map is also shown in Figure~\ref{c18o-mom}.
We see a clear enhancement in the velocity dispersion toward the
extinction ring. Moreover, the morphology of the velocity dispersion
and the \av\, appear to match very well. Indeed, of all the measured
quantities, the velocity dispersion matches best with the visual
extinction.

In addition to the velocity gradient, the line profiles clearly show
an asymmetry which changes across the Globule. To quantify the
asymmetry of the emission profiles we have calculated the `skewness'
of the emission at each position within the map. We follow
\cite{Gregersen97} and define skewness as a dimensionless ratio of the
third moment to the 3/2 power of the second moment where both are
normalized by the first moment, such that

\[ {\rm {skewness}} = \frac{\int T_{mb}(v) (v-v_{LSR})^{3}\,dv}{\int T_{mb}(v)\,dv} / \left [ \frac{\int T_{mb}(v) (v-v_{LSR})^{2}\,dv}{\int T_{mb}(v)\,dv} \right ]^{3/2} \]

\noindent where \vlsr\, is the mean velocity of the Globule
($\sim$$-$5.84\,\kms). Figure~\ref{c18o-mom} also shows the skewness
image and reveals that, toward the west of the extinction ring, the
profiles show a negative (blue) asymmetry, while toward the east, the
profiles typically show a positive (red) asymmetry.

\subsection{Temperature and density tracers}

In addition to the \ceo\, map, we have also obtained \co\, and CS maps
toward the Globule. While these results will be discussed in detail in
Saul et al. (in prep), we include them here to help with the
interpretation of the \ceo\, results. Because \co\, is typically
optically thick, it can be used to trace the kinetic temperature of
the gas. Because it has a large dipole moment, CS has a high critical
density, and is typically excited collisionally only in high-density
gas ($n {\gtrsim} 10^5$ cm$^{-3}$). Thus, the combination of \co\, and
CS can trace the warm, dense gas within the Globule.

\begin{figure}
\begin{center}
\includegraphics[angle=-90,width=0.5\textwidth]{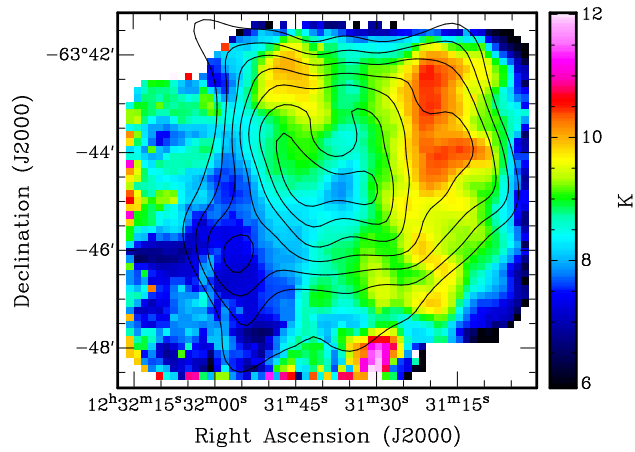}
\caption{\label{temp}Kinetic temperature (\tk) across the Globule 
        derived from the optically thick \co\, emission. Overlaid are
        contours of the smoothed visual extinction image (levels are
        from 6 to 12 in steps of 0.75 magnitudes). We see a
        temperature gradient across the Globule of 9.6 $\pm$ 0.5\,K in
        the north-west to $\sim$ 7.6 $\pm$ 0.4\,K in the south-east.}
\end{center}
\end{figure}

Figure~\ref{temp} shows the kinetic temperature map overlaid with
contours of the smoothed visual extinction map. We find that the
kinetic temperature displays a range of values over the Globule (from
$\sim$ 8 to 10\,K). The highest temperatures are not associated with
the extinction ring, but are located to the north-west (\tk\, of 9.6
$\pm$ 0.5\,K compared to $\sim$ 7.6 $\pm$ 0.4\,K in the
south-east). The kinetic temperature of the gas associated with the
extinction ring is $\sim$ 8.8 $\pm$ 0.4\,K.

\begin{figure}
\begin{center}
\includegraphics[angle=-90,width=0.4\textwidth]{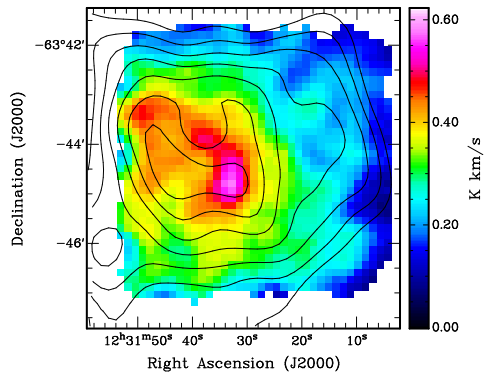}
\caption{\label{cs} CS integrated intensity image overlaid with contours 
    of the smoothed visual extinction image.  The contour levels are
    from 6 to 12 in steps of 0.75 magnitudes. Bright CS emission is
    seen coincident with the extinction ring, however, their overall
    morphologies are different. The morphology of the CS integrated
    intensity emission matches very well the \ceo\, integrated
    intensity.}
\end{center}
\end{figure}

We find that the dense gas (as traced by the CS integrated intensity)
is associated with the extinction ring (Fig.~\ref{cs}). While the
morphology of the CS emission and extinction do not match exactly,
there is a general correspondence between the two. However, the
morphology of the CS emission agrees very well with the \ceo\,
emission: both peak on the western arc of the extinction ring. The
fact that the \ceo\, and CS emission match so well suggests that,
chemically, this core does not appear to be evolved. Because
carbon-chain molecules are considered `early-time' molecules, they are
abundant in chemically young cores and depleted in more evolved,
star-forming cores. This is in contrast to molecules such as \nnhp\,
and \ammonia, which are considered `late-time' molecules and, thus,
are more abundant in chemically evolved cores
\citep{Bergin97,Aikawa05,Suzuki92}.

\section{Discussion}

\subsection{Comparison of the dust extinction to molecular abundances}

We find that the morphology of the \ceol\, integrated intensity
generally matches the visual extinction. By smoothing the \av\, image
to match the angular resolution of the \ceo\, map, we can make a
point-by-point comparison between the integrated
\ceo\, emission and the visual extinction across the Globule. 

\begin{figure}
\begin{center}
\includegraphics[angle=90,width=0.45\textwidth]{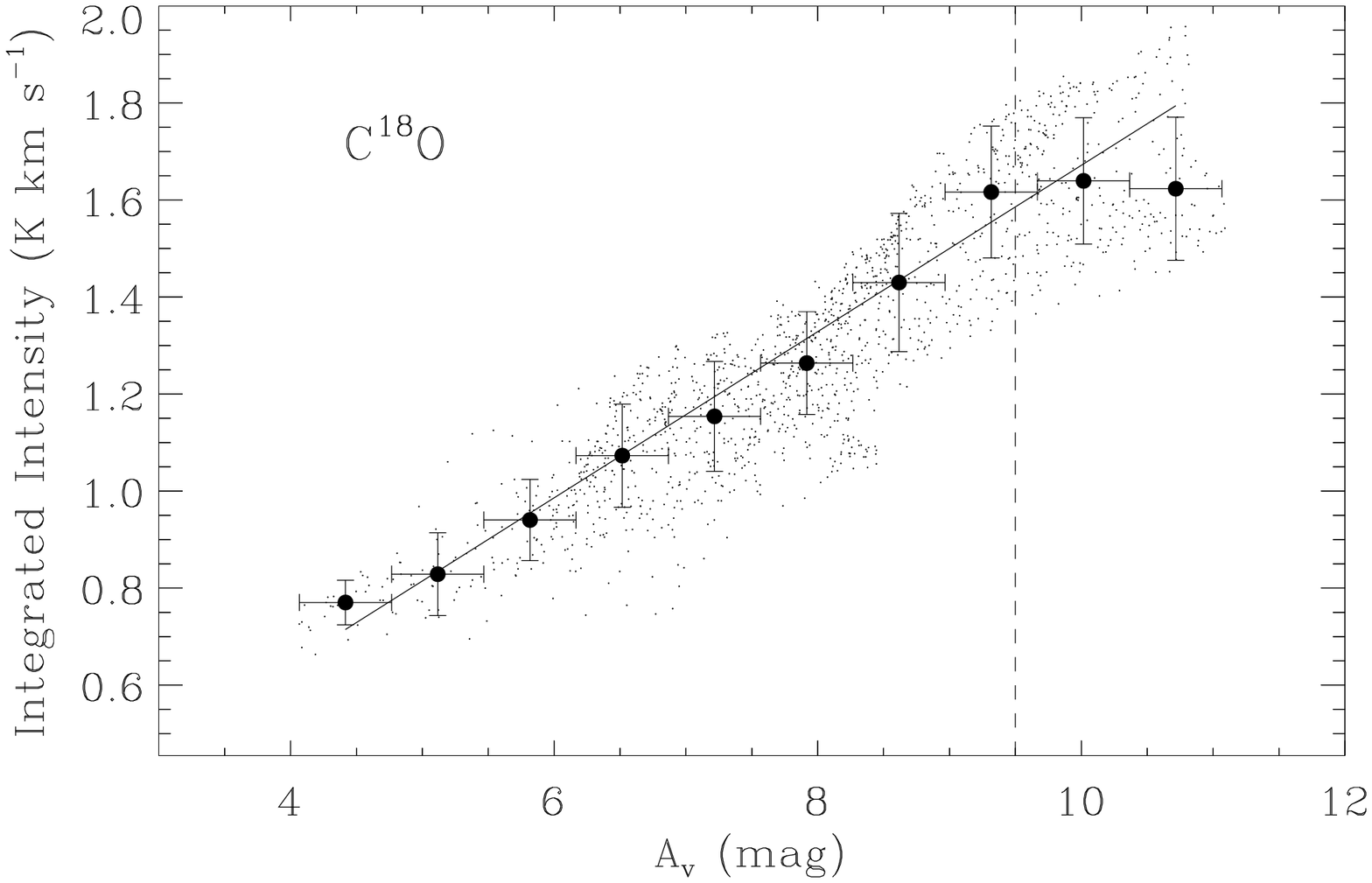}
\includegraphics[angle=90,width=0.45\textwidth]{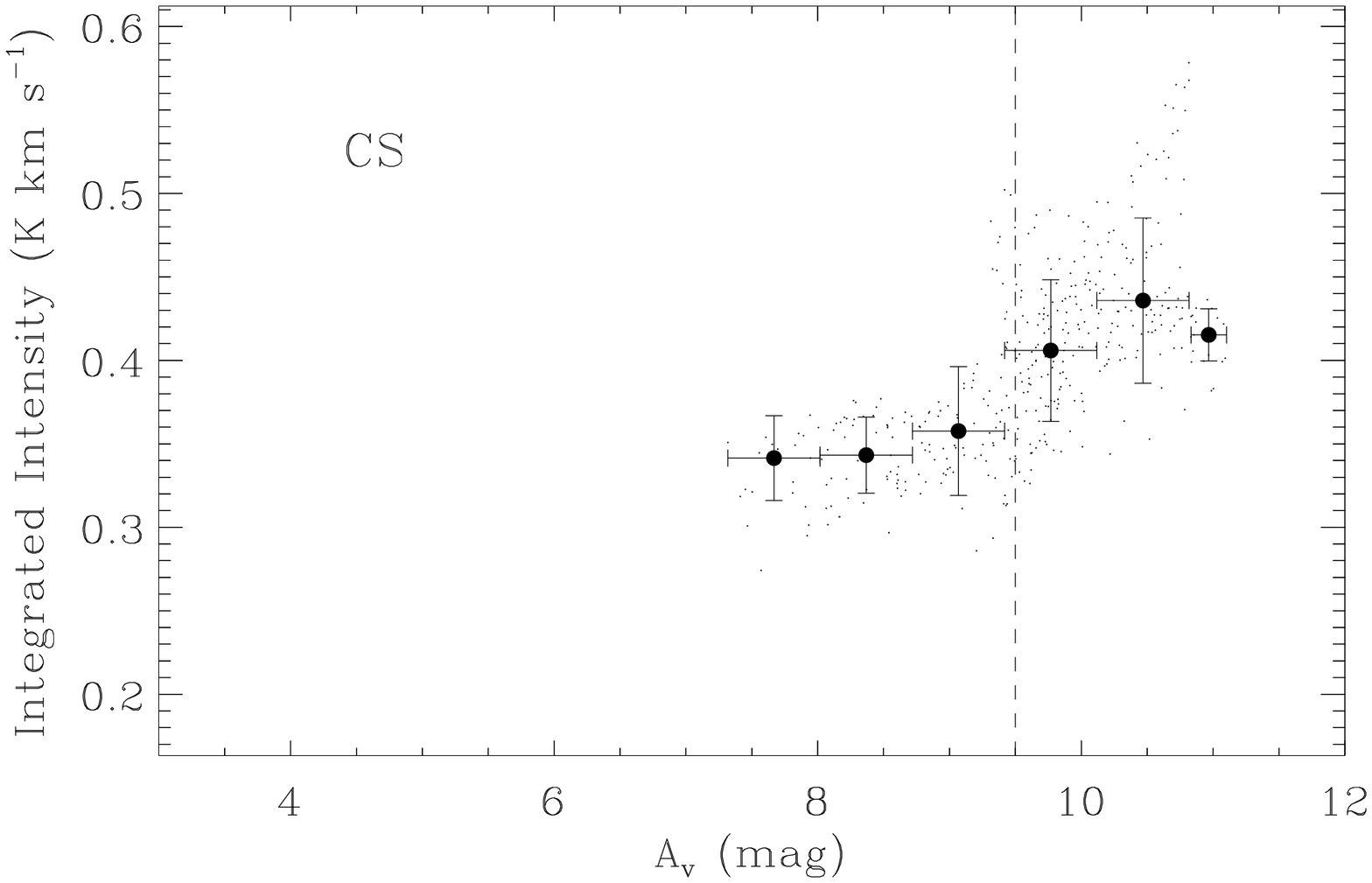}\\
\caption{\label{c18o-av}\ceo\, (left) and CS (right) integrated intensity as a function of 
       visual extinction (\av) for all positions across the
       Globule. The individual data points are shown as small dots
       while the filled circles were generated by taking the median of
       the integrated intensities within evenly-spaced \av\, bins (0.7
       mags).  The errors bars in \av\, represent the range in each
       bin. The error bars in the integrated intensity represent the
       dispersion in the data.  The dashed vertical line marks the
       approximate visual extinction at the edge of the extinction
       ring ($\sim$ 9.5 mags).  The solid line is a least-squares fit
       to the data for 4 mags $<$ \av\, $<$ 9.5 mags. We see a general
       correlation between the \ceo\, and CS integrated intensity and \av\,
       over the complete range in \av, however, we also note that both the
       \ceo\, and CS appear depleted at high extinctions (\av\,$>$ 9.5
       mags).}
\end{center}
\end{figure}

Figure~\ref{c18o-av} shows the \ceo\, integrated intensity as a
function of \av. The individual data points are shown as small dots
while the filled circles trace the overall trend and were generated by
taking the median of the integrated intensities within evenly-spaced
\av\, bins.  Each \av\, bin is 0.7 mag wide; the errors bars in
\av\, represent the range in each bin. The error bars in the 
integrated intensity represent the dispersion in the data.  We find
that the \ceo\, integrated intensity is generally correlated with
visual extinction across the complete range of measured extinction
values.  However, for any given \av, there appears to be a range in
the measured integrated intensities. Similar to what is observed
toward other dark clouds, we find that \ceo\, appears to be
depleted only at the highest measured extinction (\av\, $\gtrsim$ 9.5 mag).
Figure~\ref{c18o-av} also shows a similar plot for the CS
integrated intensity as a function of \av\, and
reveals that CS is also only depleted at the very highest extinctions.  A
similar analysis of the \ceo\, integrated intensity as a function of
\av\, toward the denser core B68 \citep{Bergin02} shows that
\ceo\, is depleted for \av\,$>$ 10 mags. Thus, we do not see significant 
molecular depletion toward the Globule primarily because its mean
column density is too low. The lack of a high density, centrally
condensed core further suggests that the Globule is relatively young.

A least-squares fit to the \ceo\, data (solid line) for
4~mag$<$\av$<$9.5~mag reveals that they are well matched to a linear
fit with a constant slope over the complete range of measured
extinction values. The derived linear fit is

\[ I(\ceo) =  (0.17 \pm 0.01) A_{v} - 0.04 \pm 0.07 {\rm {\hspace{0.2cm} K\,\, km\,s^{-1}.}} \]

The relative abundance of \ceo\, to molecular hydrogen (\hh) can be
estimated from comparisons between the derived values of \ceo\, column
density, N(\ceo), and \av. A least-squares linear fit to these data
also show a linear correlation which is given by

\[N(\ceo) =  (1.4 \pm 0.1) \times 10^{14} A_{v} + (0.8 \pm  0.4) \times 10^{14} {\rm {\hspace{0.2cm} cm^{-2}}} \]

Our derived slope is slightly lower than that calculated by
\cite{Lada94} within IC 5146 ($\sim$2.3 $\times 10^{14}$ \cms) but is
consistent with the range observed toward other dark clouds (0.7--3.5
$\times$ 10$^{14}$ \cms\, mag$^{-1}$; see
\citealp{Hayakawa99,Harjunpaa04} and references therein). The tight
correlation in the integrated intensity and, hence, \ceo\, column
density over all measured \av, suggests that there is little variation
in the molecular abundance of \ceo\, as a function of \av\, and that
the assumption of a single excitation temperature across the Globule
is most likely valid.

By converting the measured \av\, to a \hh\, column density, where
N(\hh)/\av =$10^{21}$~\cms\,mag$^{-1}$, we can also derive the
relation between the \ceo\, and \hh\, column densities. A
least-squares fit to plots of the data measured as N(\ceo) and N(\hh)
gives the relation

\[ N(\ceo) = (1.4 \pm 0.6)\times 10^{-7}\, N(H_{2}) - (0.8 \pm 0.4) \times 10^{14}  {\rm {\hspace{0.2cm} cm^{-2}}} \]

\noindent The derived slope is comparable to the value of $\sim 1.8 \times 10^{-7}$ 
calculated by \cite{Frerking82} toward the dense cores within Taurus
and $\rho$ Ophiuchus and to the values determined for the dense cores
L~977 and FeSt 1-457 \citep{Alves99,Aguti07}.

\begin{figure}
\begin{center}
\includegraphics[angle=90,width=0.5\textwidth]{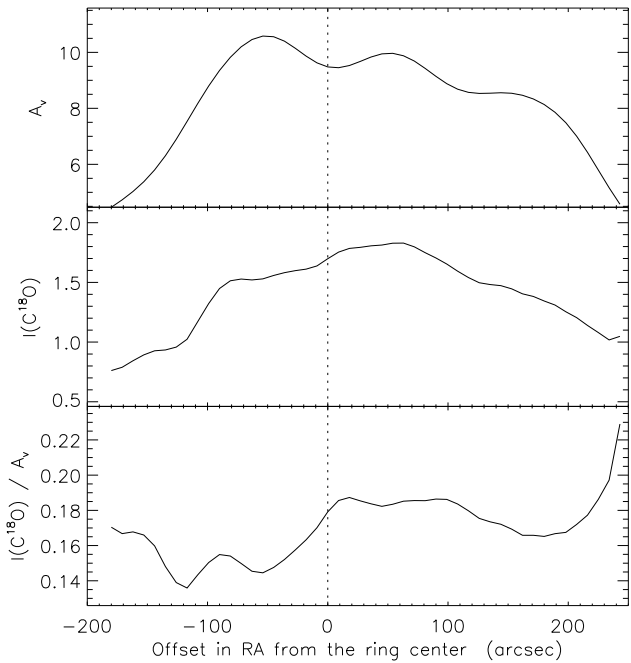}
\caption{\label{av_ii_ratio}Visual extinction (\av), \ceo\, 
     integrated intensity, and the ratio of \ceo\, integrated
     intensity to \av\, across the Globule. These plots were generated
     by taking the mean values in a 45\,\arcsec\, strip of constant
     Declination across the Globule. The \av\, across the Globule
     shows the enhanced extinction toward the ring and its central
     10\% decrement. The \ceo\, integrated intensity does not show a
     similar decrement. The lower panel shows that the ratio is
     essentially constant toward the western arc of the ring (positive
     offsets), but drops by $\sim$ 20\% toward the eastern arc
     (negative offsets).}
\end{center}
\end{figure}

Even though we see very little overall \ceo\, depletion toward the
Globule, the apparent differences between the peaks of the visual
extinction and \ceo\, in the ring may arise as a result of
depletion. Because the western arc of the ring has a higher mean
extinction compared to the eastern arc, it is possible that the \ceo\,
is more depleted toward the western arc.  These difference are seen in
Figure~\ref{av_ii_ratio} which plots the mean extinction, \ceo\,
integrated intensity, and the ratio of \ceo\, integrated intensity to
\av\, across the Globule. For this analysis, the \av\, image was 
smoothed to match the angular resolution of the \ceo.  These plots
were generated by taking the mean values in a 45\arcsec\, strip of
constant Declination across the Globule.  The enhanced extinction
toward the ring is clearly seen, along with its central hole which is
evident as an $\sim$ 10\%\, decrement in the column density (top
panel). This is in contrast to the \ceo\, integrated intensity (middle
panel) which does not show a significant decrement. These differences
are most obvious in the ratio of the \ceo\, integrated intensity to
\av\, (lower panel). We find that the ratio is essentially constant
toward the western arc of the ring (positive offsets), but drops by
$\sim$ 20\% toward the eastern arc (negative offsets). This is
consistent with the fact that depletion is typically only seen toward
regions with the highest \av\, ($>$ 10 mags; e.g.,
\citealp{Bergin02}).  Because the mean \av\, within most of the
extinction ring is very close to this threshold, it may be that we are
seeing the early onset of detectable \ceo\, depletion in the western
arc of the extinction ring.  It is possible that more depletion is
seen here simply because it has a higher column density.

This is consistent with a recent study of the solid-state H$_{2}$O ice
column densities toward the Globules within the Coalsack
\citep{Rodgers07}. Toward the outer edges of Globule 2 they find only a small amount of oxygen
depletion. Because the time required to accrete the observed amount of
oxygen is significantly shorter than the time for full oxygen
depletion, they conclude that the Globule is relatively young
($\lesssim$ 10$^{5}$ years; \citealp{Rodgers07}). The fact that this
depletion time scale is comparable to the Globule's dynamical time
scale \citep{Lada04} and the small amount of detectable depletion
confirms the youth of the Globule.


\subsection{Global properties of the Globule}

While Globule 2 is one of the densest cores within the Coalsack
Nebula, with a mean density n(\hh) of $\sim$2.7$\times$10$^{3}$\,\cmc,
it is less dense than typical cores found in Taurus
($\sim$1.2$\times$10$^{4}$\cmc; \citealp{Ladd00}) and the Pipe Nebula
($\sim$7.3$\times$10$^{3}$\,\cmc; \citealp{Lada08}). Because it is
quiescent and does not harbor a central protostar, it is considered a
dense, starless core. Moreover, the high Jeans mass to total mass
ratio and the Bonnor-Ebert fit to the density profile for the outer
regions of the Globule, suggest that the Globule is stable against
collapse \citep{Lada04}.  However, the discovery of an extinction ring
within the Globule has lead to speculations that it may be a
dynamically unstable, transient core on the verge of condensing to
form a protostar. With our \ceo\, map, we can now test this idea and
determine whether the Globule is gravitationally bound, in virial
equilibrium, magnetically super-critical, rotating, thermally
dominated, and/or pressure confined. These properties can reveal
important clues to the nature and evolutionary status of the Globule.

\subsubsection{Gravitationally bound?}

From the single line profile toward Globule 2, \cite{Lada04} suggested
that the Globule is gravitationally bound.  We now extend this
analysis to include all the data measured toward the Globule. By
comparing the escape velocity, $V_{esc}$, to the average measured
three-dimensional velocity dispersion across the Globule, we can
determine if the Globule is gravitationally bound.  The escape
velocity is given by

\[ V_{esc} = \sqrt{2GM/R} \]

\noindent where M and R are the Globule's mass and radius respectively. 
We use the values estimated from the dust extinction for the mass
(M=6.1\,\Msun) and the radius calculated from the Bonnor-Ebert fit
(R=290\arcsec, 0.21 pc, or 6.51$\times$10$^{13}$ cm;
\citealp{Lada04}).  Given these values, we estimate that the V$_{esc}$
for Globule 2 is 0.50\,\kms.

The average three-dimensional velocity dispersion, $\sigma_{3d}$, is
calculated via the expression

\[ \sigma_{3d} = \sqrt{3a^{2} + 3\sigma_{nt}^{2}} \]

\noindent where $a$ is the one-dimensional isothermal sound speed 
(0.19\,\kms\, for hydrogen in a 10~K gas) and $\sigma_{nt}$ is the
one-dimensional non-thermal velocity dispersion, i.e. $\sigma_{nt} =
\sqrt{\sigma^{2} - \sigma_{th}^{2}}$. The measured velocity dispersion, 
$\sigma$, was estimated from the average in the second moment map
across the Globule and is $\sim$ 0.18\,\kms. Assuming the gas is at
10~K, the one-dimensional thermal velocity dispersion, $\sigma_{th}$,
is 0.05\,\kms.  Thus, the non-thermal velocity dispersion,
$\sigma_{nt}$, is $\sim$ 0.17\,\kms\, and implies that the turbulent
motions are subsonic. Using these values we find that the average
three-dimensional velocity dispersion, $\sigma_{3d}$, is
0.44\,\kms. Because $\sigma_{3d}$ is formally less than the escape
velocity, it is likely that the Globule is indeed gravitationally
bound, although only marginally so.

The one-dimensional virial velocity dispersion, $\sigma_{virial}$, was
also calculated via the expression

\[ \sigma_{virial} = \sqrt{\frac{1}{5}\frac{GM}{R}} \]

\noindent and was found to be 0.16\,\kms. This is lower  than both 
the sound speed in a 10 K gas (0.19\,\kms) and the measured
\ceo\, velocity dispersion across the Globule (0.18\,\kms). The fact
that the $\sigma_{virial}$ is lower than the measured velocity
dispersions indicates that the Globule is not virialized.  Thus, it
appears that the Globule is only marginally bound and not in virial
equilibrium, implying that is it extremely young.

\subsubsection{Magnetically super-critical?}

The role of magnetic support within the Globule can be measured via
the ratio of the observed mass to magnetic flux ($\Phi$) represented
in terms of the critical value. We follow \cite{Crutcher07} and define
this ratio as

\[ \lambda =\frac{(M/\Phi)_{observed}}{(M/\Phi)_{critical}} = 7.6 \times 10^{-21} \frac{N(H_{2})}{B}\]

\noindent where N(\hh) is in \cms\, and B is in $\mu$G. To calculate the magnetic field, B,
we use the \cite{CF53} equation 

\[ B = \frac{\sigma_{v}}{\sigma_{\Theta}} \sqrt{4 \pi \rho} = 10 \sqrt{\frac{n(H_{2})}{2.7 \times 10^{3}}} \mu G\]

\noindent where $\sigma_{v}$ is the one-dimensional  velocity dispersion and 
$\sigma_{\Theta}$ is the dispersion in the measured polarization
angles. Using $\sigma_{v}$=0.18\,\kms, $\sigma_{\Theta}$=0.66 rad
\citep{Jones84}, and a molecular hydrogen density, n(\hh), of
2.7$\times$ 10$^{3}$\,\cmc, we find that B$\sim$10$\,\mu$G. Thus, the
directly observed ratio, $\lambda_{observed}$, is $\sim$ 4.8. Applying
corrections for geometrical effects of 1/3 \citep{Crutcher07}, we find
that $\lambda$ is $\sim$ 1.4. A value of $\lambda$ greater than unity
implies that the magnetic field is super-critical. Thus, we find that
the Globule is marginally super-critical which suggests that the
Globule is not dominated by magnetic support and that it is likely
bound if not already in the process of gravitational contraction.

\subsubsection{Rotating?}

As mentioned previously, the channel maps, ($\ell$-$V$) diagram, and
first moment map (the intensity weighted velocity field) all show
signs of a velocity gradient across the Globule. In the \ceo\, first
moment map (Fig.~\ref{c18o-mom}), the lines of constant velocity are
nearly parallel and are aligned north-south. If the velocity gradient
is linear when projected on the sky, we can relate the observed
velocity of any point on the projected surface of the Globule to the
velocity gradient via the expression \citep{Goodman93,Lada03a}

\[ v_{LSR} = v_{0}  + \frac{dv}{ds}\,\Delta \alpha\, {\rm {cos}} \theta + \frac{dv}{ds}\,\Delta \delta\,{\rm {sin}}\theta \]

\noindent where $\Delta \alpha$ and $\Delta \delta$ are the offsets in Right
Ascension and Declination (in arcsec), $dv/ds$ is the magnitude of the
velocity gradient in the plane of the sky, $\theta$ is its direction
(measured from east of north), and $v_{0}$ is the systemic
radial velocity of the Globule. The magnitude and direction of the
velocity gradient and the Globule's systemic velocity were determined
through a least-squares fit to the observed \ceo\, velocity
distribution (as shown in Fig.~\ref{c18o-mom}). We find that $dv/ds$ =
1.00 $\pm$ 0.01~\kms\,pc$^{-1}$, $\theta$ = 90.1 $\pm$ 0.3\degree, and
$v_{0}$ = $-$5.81 $\pm$ 0.01\,\kms. This velocity gradient is very
similar to that found in the \ceo\, survey of the Coalsack Nebula by
\cite{Kato99}. Even though these data were under-sampled (2\farcm7
beam on a 4\arcmin\, grid), they reveal a velocity gradient across
Globule 2 of $\sim$ 0.90\,\kms\,pc$^{-1}$. If the Globule was
undergoing solid-body rotation whose axis is perpendicular to the
line-of-sight, then this velocity gradient would correspond to an
angular velocity of 2.9 $\times$ 10$^{-14}$ s$^{-1}$ \citep{Kato99}. 

It is possible that the velocity gradient arises due to the rotation
of the Globule. Assuming solid-body rotation, the ratio of rotational
kinetic energy to gravitational potential energy, $\beta$, can be used
to determine the significance of rotation to the overall dynamics of
the Globule. Following \cite{Goodman93}, we calculate $\beta$ via the
expression

\[ \beta = \frac{(1/2) I w^{2}}{qGM^{2}/R} = \frac{1}{2}\frac{p}{q}\frac{w^{2}R^{3}}{GM}\]

\noindent where $M$ and $R$ are the mass and radius of the Globule 
respectively, the moment of inertia is given by $I=pMR^{2}$, $q$ is
defined such that the gravitational potential energy is $qGM^{2}/R$,
and $w=(dv/ds)/{\rm {sin}}\,i$. We assume a sphere with a r$^{-2}$
density profile, which implies $p/q$ is 0.22, and that sin\,$i$=1.
Using M=6.1\,\Msun, R=6.51$\times$10$^{17}$ cm, and
$dv/ds$=1.00\,\kms\,pc$^{-1}$, we find that $\beta$=0.04. This value
of $\beta$ is similar to other cores: for example,
\cite{Goodman93} find that most clouds have $\beta \le$ 0.05.

Thus, we calculate that the apparent rotation does not contribute
significantly to the overall stability of Globule 2. In any event, it
appears that the observed dynamics of the Globule are more complicated
than simply solid body rotation and it is likely, therefore, that the
observed velocity gradient is not tracing a coherent rotating
structure. Nevertheless, the energy in the velocity gradient is
unimportant relative to gravity.



\subsubsection{Thermally dominated?}


The one-dimensional \ceo\, velocity dispersions ($\sigma \sim$
0.18\,\kms) are $\sim$ 3.6 times broader than what would be expected
if the profiles were purely dominated by thermal motions. This
suggests that significant non-thermal motions characterize the
velocity field of the gas and that turbulence may play a role in the
internal pressure of the Globule.  To quantify the degree to
which turbulence may be contributing to the internal pressure,
we calculate the ratio of the thermal to non-thermal (turbulent)
pressure as given by

\[ R_{p} = \frac{a^{2}}{\sigma_{nt}^{2}}\]

\noindent where $a$ is the one-dimensional isothermal sound speed 
(0.19\,\kms\, for hydrogen in a 10~K gas) and $\sigma_{nt}$ is the
one-dimensional non-thermal velocity dispersion. Taking the average
$\sigma_{nt}$ across the Globule, we estimate the ratio of thermal to
non-thermal pressure, R$_{p}$, is $\sim$~1.2. Thus, it appears that
thermal pressure is comparable to, or slightly greater than, the
non-thermal, or turbulent, pressure within the Globule.

While it is useful to calculate average values across the Globule,
there are clear differences in the \ceo\, emission within and either
side of the extinction ring. Indeed, if we take the typical measured
velocity dispersion across the extinction ring and compare it to two
regions on opposing sides of the ring, we find that the velocity
dispersion toward the ring ($\sigma$\, of 0.20 $\pm$ 0.01\,\kms) is $\sim$
1.3 times greater than that observed outside the ring (0.15$\pm$ 0.01
and 0.17$\pm$ 0.01\,\kms\, to the east and west of the ring
respectively). Using these values we find that the non-thermal
velocity dispersions ($\sigma_{nt}$) in and out of the ring are 0.19
and $\sim$0.15\,\kms\, respectively. Thus, we see turbulent sonic
motions within the ring and turbulent subsonic motions external to it.

The differences between these regions are most obvious when we
consider the ratio of the thermal to turbulent pressures, $R_{p}$:
within the ring, $R_{p}$ is $\sim$ 1, while external to the ring
$R_{p}$ is $\sim$1.6. This suggests that thermal pressure is more
dominant external to the ring, whereas toward the ring, the
contributions from thermal and turbulent pressure are comparable.

\begin{figure}
\begin{center}
\includegraphics[angle=90,width=0.45\textwidth]{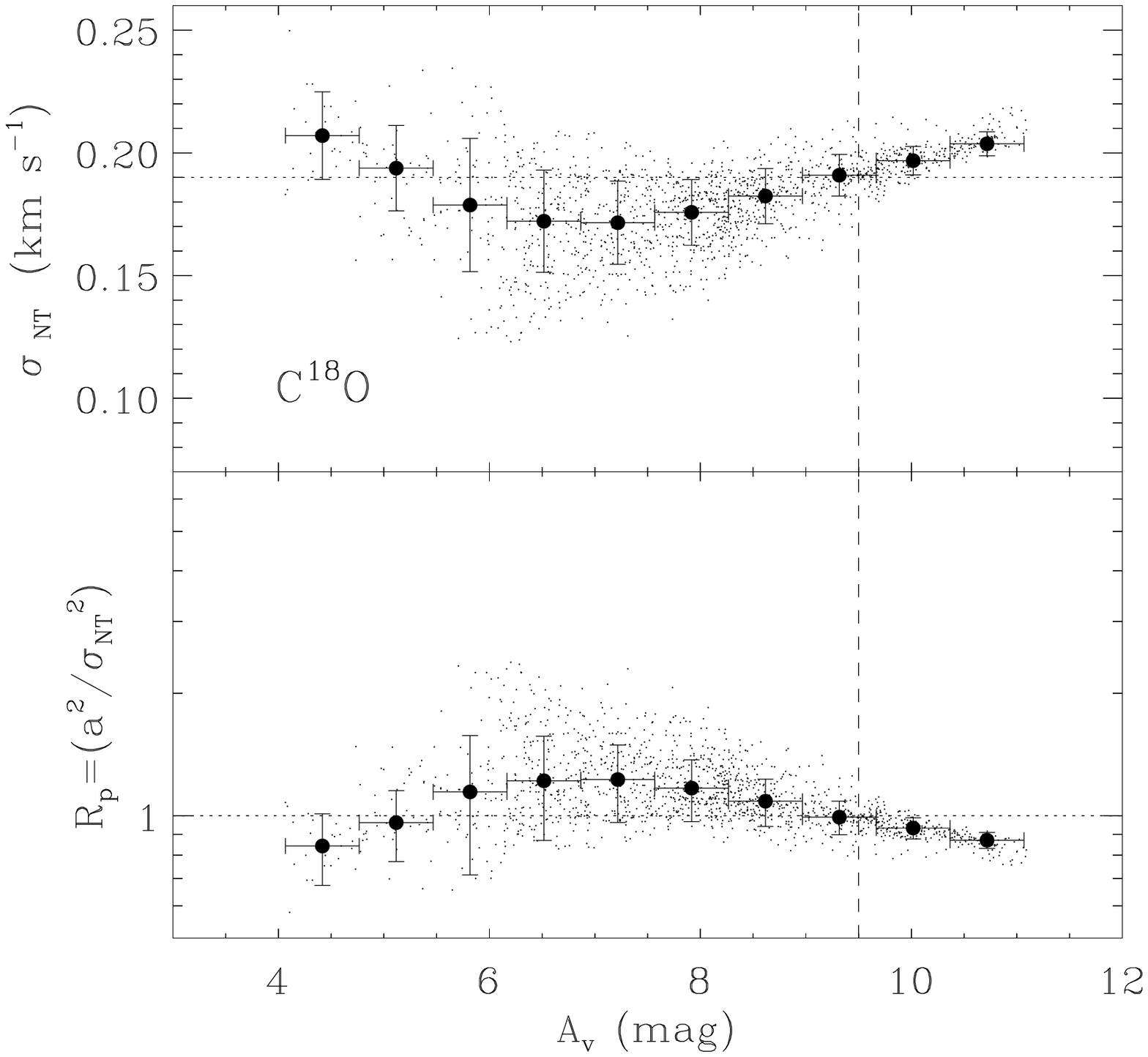}
\includegraphics[angle=90,width=0.45\textwidth]{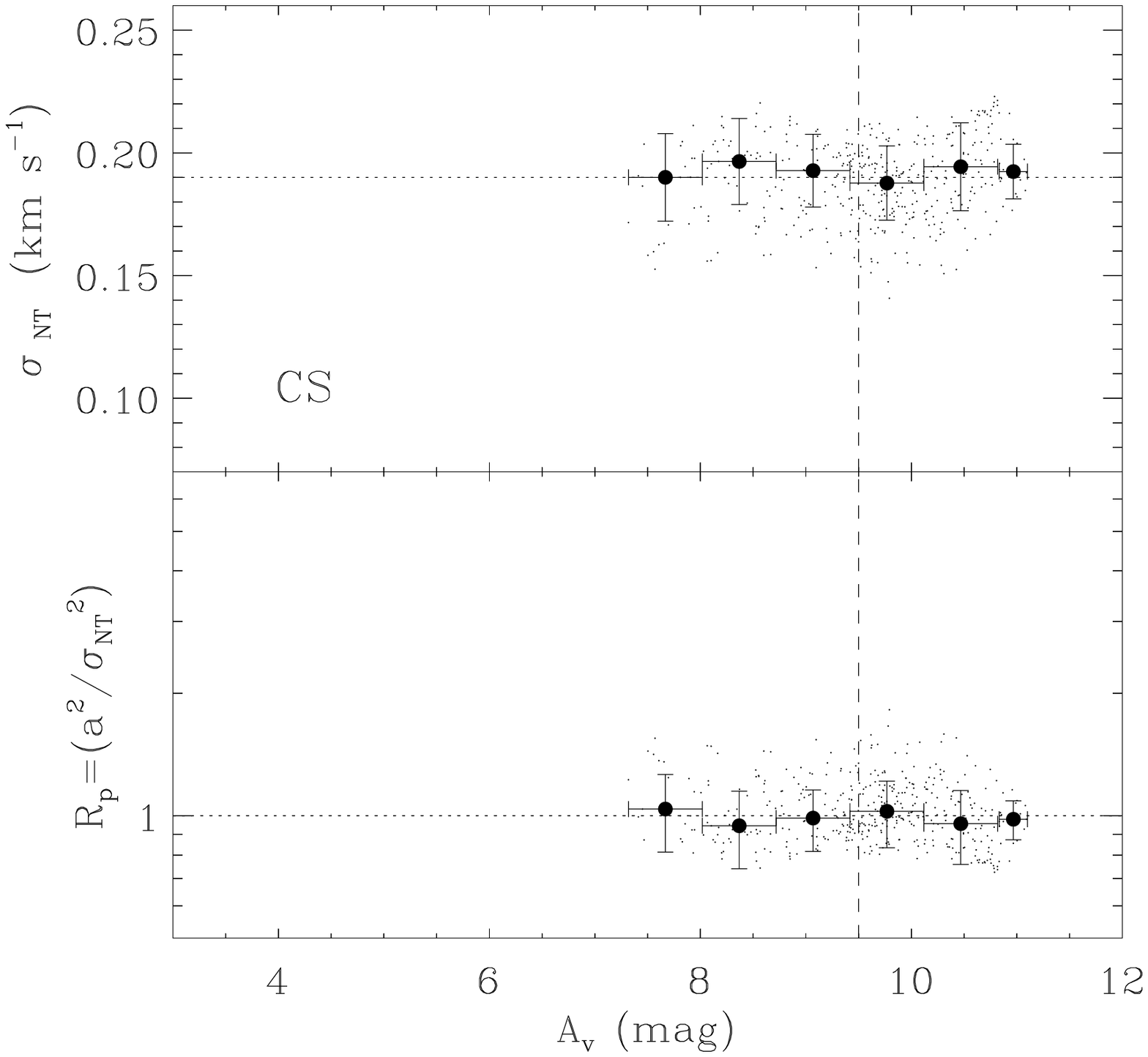}
\caption{\label{c18o-av_mom2}Measured non-thermal 
   velocity dispersion ($\sigma_{nt}$; upper panels) and the ratio of
   thermal to non-thermal pressures (R$_{p}$; lower panels) as a
   function of \av\, for all positions with the Globule (\ceo\, left,
   CS right).  The individual data points are shown as small dots
   while the filled circles were generated by taking the median of the
   integrated intensities within evenly-spaced \av\, bins.  Each \av\,
   bin is 0.7 mag wide; the errors bars in \av\, represent the range
   in each bin. The error bars in $\sigma_{nt}$ and R$_{p}$ represent
   the dispersion in the data. In the upper panels the dotted
   horizontal line marks the one-dimensional isothermal sound speed
   (0.19\,\kms\, for hydrogen in a 10 K gas). In the lower panels the
   dotted horizontal line marks the point at which thermal and
   turbulent pressure are equal.  The dashed vertical line marks the
   approximate visual extinction at the edge of the extinction ring
   ($\sim$ 9.5 mags).  For \ceo, we see that as \av\, increases the
   non-thermal velocity dispersion increases while the ratio of
   thermal to non-thermal pressure decreases. Coincident with the ring
   $\sigma_{nt}$ transitions from subsonic to trans-sonic and R$_{p}$
   transitions from greater than to less than 1.  This suggests that
   thermal pressure dominates external to the extinction ring and that
   turbulence may be more important within the ring. In contrast,
   $\sigma_{nt}$ and R$_{p}$ appear constant across the Globule as
   traced by the CS emission.}
\end{center}
\end{figure}

This can also be seen in the left panels of Figure~\ref{c18o-av_mom2}
which shows the \ceo\, non-thermal velocity dispersion ($\sigma_{nt}$)
and the ratio of thermal to non-thermal pressure (R$_{p}$) as a
function of \av\, for all positions in the map. In the upper panel the
dotted horizontal line marks the one-dimensional isothermal sound
speed. In the lower panel the dotted horizontal line marks the point
at which thermal and turbulent pressure support are equal. The dashed
vertical line marks the approximate visual extinction at the edge of
the extinction ring ($\sim$ 9.5 mags). These plots show that the
general trend is for higher non-thermal velocity dispersions,
$\sigma_{nt}$, and lower values for R$_{p}$ as \av\,
increases. Coincident with the extinction ring $\sigma_{nt}$
transitions from subsonic to trans-sonic and R$_{p}$ transitions from
greater than to less than 1.  Thus, it appears that the outer regions
of the Globule are thermally dominated, while toward the extinction
ring non-thermal pressure may be more important.

\subsubsection{Pressure confined?}

To further characterize its nature, we have also calculated the
internal gas pressures within the Globule. The total pressure within
the Globule is measured by a combination of the pressures due to
thermal (P$_{th}$) and non-thermal motions (P$_{nt}$). The thermal and
non-thermal pressures were calculated via the expressions
\[ P_{th} = \rho a^{2} \hspace{0.5cm} {\mathrm {and}}  \hspace{0.5cm} P_{nt} = \rho \sigma_{nt}^{2} \]
\noindent where $\rho$ is the mean density, $\rho = 3M/(4\pi R^{3})$, 
$a$ is the one-dimensional isothermal sound speed, and $\sigma_{nt}$
is the one-dimensional non-thermal velocity dispersion. We find that
P$_{th}/k \sim 2.8\times 10^{4}$~K~\cmc\, and P$_{nt}/k \sim 2.2\times
10^{4}$~K~\cmc. Thus, the total pressure, P$_{total}/k$, due to
thermal and non-thermal motions is $\sim$ 5.0$\times 10^{4}$\,K\,\cmc. 
%
%
%
%

Because the Globule lies within the larger Coalsack molecular cloud
complex, it is possible that the weight of the Coalsack molecular
cloud may also be a significant source of confining pressure. The
pressure due to the weight of the cloud, P$_{cloud}$, is given by
\[P_{cloud} = \frac{3 \pi}{20} G \Sigma^2 \phi_{G} \]
\noindent where $G$ is the Gravitational constant, $\Sigma$ is the 
mean mass surface density of the cloud ($\Sigma = M/\pi R^2$), and
$\phi_{G}$ is a dimensionless correction factor to account for the
non-spherical geometry of a cloud \citep{Bertoldi92}. Given the
Coalsack molecular cloud has a mass of $\sim$3550\,\Msun, radius of
$\sim$ 7.1 pc \citep{Nyman89}, and assuming $\phi_{G}$ = 1.6, we find
that P$_{cloud} = 0.8\times 10^{4}$~K~\cmc. 

This pressure is more than a factor of 5 lower than the internal
pressure calculated from a combination of the thermal and non-thermal
motions which suggests that it is unlikely that the weight of the
Coalsack molecular cloud contributes significantly to the overall
confining pressure of the Globule. Interestingly, this pressure is
comparable to the pressure resulting from the interstellar medium,
P$_{ISM}/ k \sim 10^{4}$\,K\,\cmc\, \citep{Bertoldi92}. Thus, we find
that the total internal pressure is greater than both P$_{ISM}/ k$ and
P$_{cloud}/ k$ and that the Globule is super-critical.  This adds to
earlier suggestions that the Globule is gravitationally bound. This is
in contrast to dense cores found within the Pipe Nebula most of which
are thought to be thermally dominated, unbound cores that are confined
by external pressure \citep{Lada08} and suggests that the Globule is
bound by gravity.

\subsection{The Origin of the Ring}

\subsubsection{External pressure driving a compression wave?}

Recent dynamical models for Globule 2 \citep{Hennebelle06} suggest
that the ring structure has formed due to a rapid increase in the
external pressure on a core with very little initial turbulence. In
this picture, the ring structure traces the dense regions swept up by
the compression wave as it propagates through the core. The central
depression exists because the compression wave has not yet converged
in the center. The models generated by \cite{Hennebelle06} predict
that toward the center of the Globule the line profiles should be
double-peaked, while toward the outer edges of the Globule the
profiles should show a single line.

\begin{figure}
\begin{center}
\includegraphics[angle=0,width=0.68\textwidth]{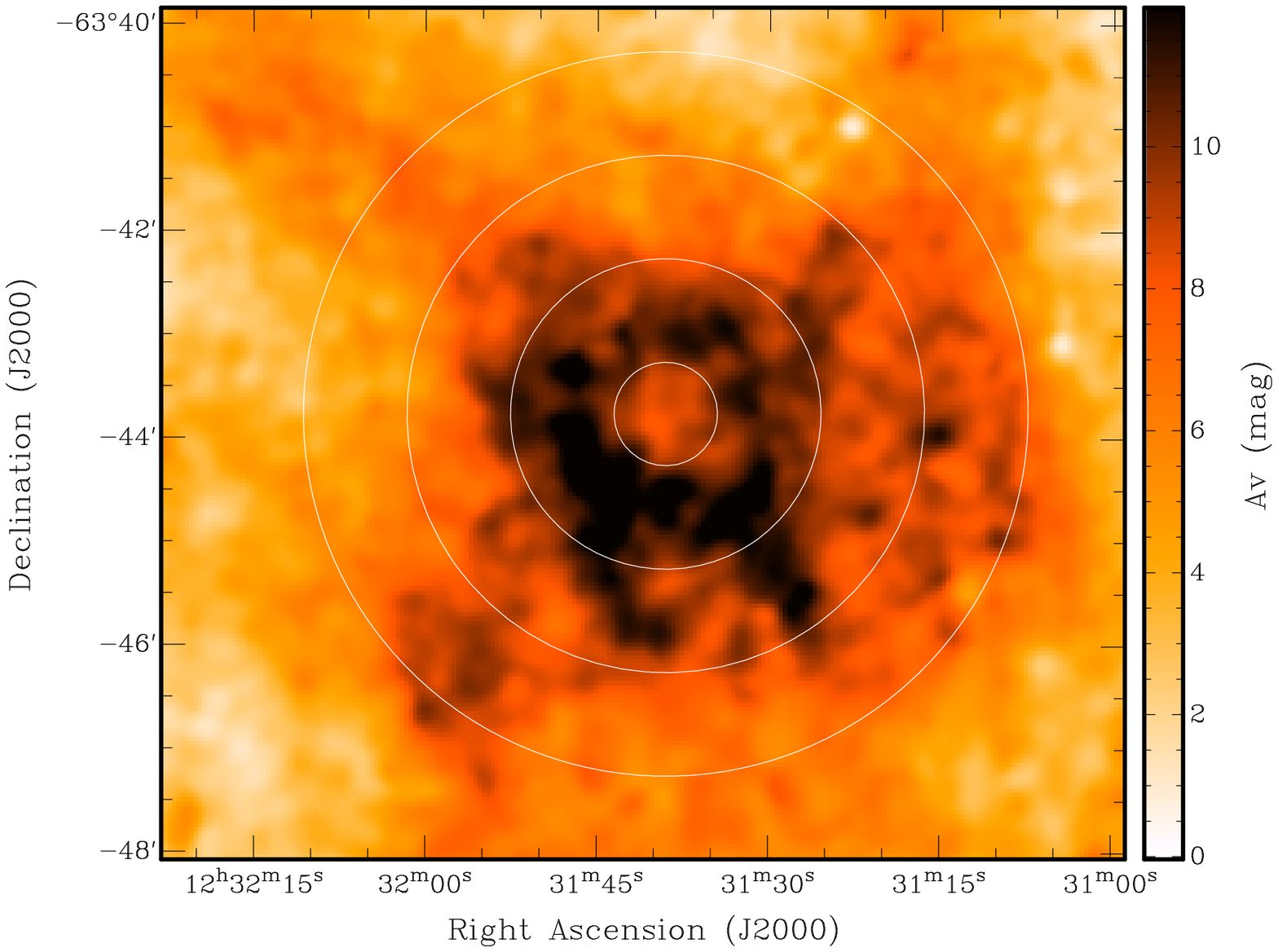}
\includegraphics[angle=0,width=0.2\textwidth]{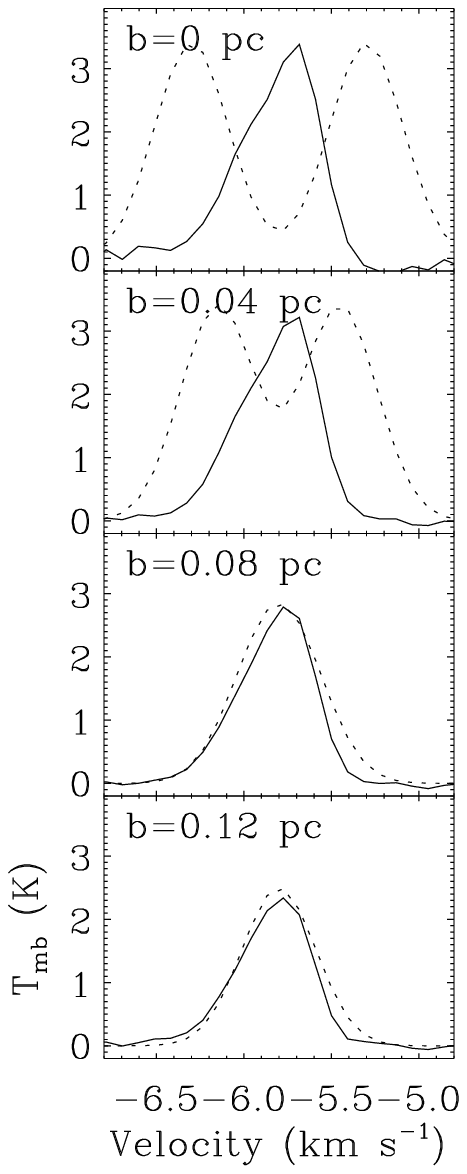}
\caption{\label{model}Comparison between our \ceo\, data and the models 
    of \cite{Hennebelle06} for different impact parameters, $b$,
    across the Globule. The spectra (right panels) were generated by
    averaging over annuli at the impact parameters $b$=0, 0.04, 0.08
    and 0.12\,pc (shown as circles on the extinction image: left
    panel). The example model spectra were estimated from figure 2 of
    \cite{Hennebelle06} and are plotted so that the center velocity
    (their V=0.0\,\kms) corresponds to the \vlsr\, of the Globule
    ($-$5.8\,\kms). The individual line centers, widths, and relative
    intensities were taken from their figure, while the peak
    temperatures were scaled to approximately match the data. We find
    that at larger impact parameters ($b >$ 0.08\,pc) the model
    matches the data well. Toward the ring ($b <$ 0.04\,pc), however,
    we see noticeable differences between the data and the model which
    might arise from the initial assumptions of the simple model
    (i.e. inflow speeds and a symmetric compression wave).}
\end{center}
\end{figure}

With our \ceo\, map we can now test this model. Due to the receding
and approaching compression wave, the \cite{Hennebelle06} model
predicts that positions within 50\arcsec\, of the center should show a
double-peaked line profile. Figure~\ref{model} shows a comparison
between our data and the model of \cite{Hennebelle06} for different
impact parameters, $b$,  across the Globule. The spectra were generated by
averaging over annuli at the impact parameters $b$=0, 0.04, 0.08 and
0.12\,pc. The example model spectra were estimated from figure 2 of
\cite{Hennebelle06} and are plotted so that the center velocity 
(their V=0.0\,\kms) corresponds to the \vlsr\, of the Globule
($-$5.8\,\kms). The individual line centers, widths and relative
intensities were taken from their figure, while the peak temperatures
were scaled to approximately match the data.

From this comparison we find that at larger impact parameters
($b$=0.08 and 0.12\,pc) the models match the data well. These regions
correspond to the edges of the Globule, off the extinction
ring. Toward the ring, however, we find the models differ considerably
from the data. While we do observe a double-line profile at impact
parameters $b <$ 0.04\,pc, the separation of the line centers and the
relative intensities of the two components are noticeably different
from what the models predict. Toward the center of the ring
($b$=0.0\,pc), we find that the two components are separated by $\sim$
0.24\,\kms, with the red component dominating. Here the models predict
the line profiles should be separated by 1\,\kms\, and have the same
relative intensity. Similarly, at an impact parameter $b$=0.04\,pc
(which completely encompasses the extinction ring), we also see a
dominant red component separated by $\sim$~0.21\,\kms\, from the
fainter blue component.

The discrepancies between the observed line profiles and the model
predictions could arise simply because of the assumptions of a
symmetric compression wave. The apparent fragmentation within, and the
non-symmetrical nature of, the extinction ring suggests that the
mechanism that has given rise to the ring is unlikely to be smooth and
symmetric.  Moreover, the predicted line separations could be higher
than that measured due to an overestimation of the speed at which the
material is imploding within the models. While the model predicts the
integrated intensity and velocity dispersion should increase within
the extinction ring, it fails to predict the velocity gradient that we
observe. It is possible that the Globule may be rotating slightly
while undergoing an implosion due to a non-symmetric increase in
external pressure. More detailed modeling with the inclusion of a
non-symmetric compression wave is necessary to test this idea.

\subsubsection{The merging of two subsonic flows?}

Although we cannot definitively rule out that the ring may have formed
via an external compression wave, the above model in its current form
is unable to completely characterize the observed kinematics within
this Globule. Because the extinction ring correlates with the \ceo\,
velocity dispersion, and, thus, appears organized, it is unlikely that
the extinction ring is simply a chance super-position of material
along the line of sight.

Rather than the implosion of the Globule due to an increase in its
external pressure, it appears that we may be seeing the shearing, or
merging, of two separate velocity features.  Gaussian fits to the
spectra across the Globule suggests there are two components that
arise within the ring; one at $\sim -$5.7\,\kms\, and the other at
$\sim -$5.9\,\kms\, (see Fig.~\ref{gauss-fits}). The emission in these
channels, shown in Figure~\ref{channels}, reveals that while both of
these components peak within the ring, they also show emission
external to, and on opposing sides of, the extinction ring. Because
both of these components have non-thermal subsonic one-dimensional
velocity dispersions, we speculate that these components correspond to
two distinct subsonic flows of turbulent gas that are merging in the
center of the Globule. 

Of all the measured parameters, we find the spatial match between the
one-dimensional velocity dispersion and the visual extinction to be
the best. Thus, the convergence of these two flows within the
extinction ring has potentially caused the enhanced column density.
It may be, therefore, that the extinction ring is neither a ring nor
shell seen in projection, but simply arises due to the merging of
molecular gas and dust as the two flows are interacting. Thus, the
increase in the velocity dispersion and non-thermal pressure toward
the extinction ring may simply reflect the fact that two flows are
interacting here.  Because the outer regions of the Globule appear
thermally dominated and stable against collapse, it may be that the
Globule is evolving from the outside in and has yet to stabilize, as
evidenced by the lack of a centrally condensed core.

If the two flows were indeed interacting within the ring, then we
would also expect to see an increase in the \ceo\, integrated
intensity toward the ring: in the absence of temperature, opacity and
depletion effects, higher velocity dispersions should translate
directly into higher integrated intensities. Although a general
increase in the integrated intensity is clearly seen toward the ring,
it is interesting to note that the \ceo\, emission and extinction do
not peak at the same location within the ring. Figures~\ref{c18o-mom}
and \ref{cs} show that the \ceo\, and CS integrated emission peaks
toward western arc of the ring whereas the extinction and velocity
dispersion peak toward the eastern arc.  These morphological
differences can be explained in terms of molecular depletion at high
extinctions. Figure~\ref{c18o-av} shows that both the \ceo\, and CS are
depleted for \av\ $>$ 9.5 mags. Because the eastern arc is dominated
by extinctions $>$ 9.5 mags it shows a higher degree of molecular
depletion compared to the western arc.

As mentioned previously, there is a temperature gradient across the
Globule that goes from a \tk\, of 9.6 $\pm$ 0.5\,K in the north-west
to $\sim$ 7.6 $\pm$ 0.4\,K in the south-east. Thus, it appears that
the two distinct velocity flows may have slightly different
temperatures. Moreover, the fact that the temperature does not
increase significantly within the ring, supports the idea that it has
not formed due to the propagation of a shock front through the
Globule. If this were the case, we would expect to see the temperature
increase in the shock front and be coincident with the ring.

\section{Conclusions}

Globule 2 is one of the densest cores within the Coalsack Nebula, but
contains no evidence for star-formation. Despite the fact that overall
it appears stable against collapse, a recent high angular resolution
extinction map reveals Globule 2 contains a central ring of enhanced
column density and, thus, may in fact be a dynamically unstable,
transient core on the verge of condensing to form a protostar.  It has
been speculated that this central ring of column density has formed
either as a result of turbulence or from a shock induced via an
external compression wave \citep{Lada04,Hennebelle06}.

Here we have combined dust extinction measurements with a high angular
resolution \ceo\, molecular line map in order to study the detailed
kinematics of Globule 2.  The \ceo\, map reveals that the Globule is
marginally gravitationally bound and not virialized.  Moreover, the
\ceo\, map reveals a complex velocity profile that appears to arise
from two distinct velocity features that are present both toward the
ring and external to it. In addition, these velocity features appear
to trace material on opposing sides of the extinction ring.

Each of these velocity flows have non-thermal subsonic velocity
dispersions.  Outside the extinction ring the Globule is thermally
dominated, however, both the measured velocity dispersion and
non-thermal pressure increase toward the extinction ring. Of all the
measured properties, we find the best spatial correspondence between
the one-dimensional velocity dispersion and the visual extinction.  We
suggest, therefore, that the extinction ring is an enhancement in
column density due to the merging of the two subsonic flows. Thus, it
appears as though we are seeing the convergence of two subsonic flows
within a bound object. Moreover, because the outer regions of the
Globule appear stable against collapse and there is no centrally
condensed core, it may be that the Globule is evolving from the
outside in and has yet to stabilize.

Finally, although we cannot entirely rule out recent dynamical models
for the formation of the extinction ring via an external compression
wave \citep{Hennebelle06}, these models are unable to completely
characterize the observed \ceo\, kinematics. The discrepancies between
the model and the data may arise because of the simplified assumption
of a spherical compression wave.

\acknowledgments

We thank the referee, Dr. Whitworth, for a thorough reading of the
paper and the useful suggestions which have improved the paper
considerably. We also thank Alyssa Goodman and Scott Schnee for
providing us with the IDL gradient fitting code.




\clearpage

\clearpage

\clearpage

\clearpage

\clearpage

\clearpage

\clearpage

\clearpage

\clearpage

\clearpage

\clearpage

\end{document}